\documentclass[12pt]{iopart}

 \expandafter\let\csname equation*\endcsname\relax
 \expandafter\let\csname endequation*\endcsname\relax
\usepackage{amsmath}
\usepackage{amssymb}
\usepackage{longtable}
 \usepackage{pdfpages}
\usepackage{graphicx}
\usepackage{array}
\usepackage{fancyhdr}

\usepackage{xcolor}
\usepackage{chemformula} 

\usepackage{hyperref}
\hypersetup{
    colorlinks=true,       
    linkcolor=blue,        
    citecolor=blue,        
    filecolor=blue,        
    urlcolor=blue,         
    runcolor=blue
}

\begin{document}

\title{First-Principles Screening of Metal-Organic Frameworks for Entangled Photon Pair Generation}

\author{Sanoj Raj$^{1,4}$, Sim\'on Paiva$^2$, Rubén A. Fritz$^2$, Felipe Herrera$^{2,3}$ and Yamil J. Colón$^{1}$}

\address{$^1$ Department of Chemical and Biomolecular Engineering, University of Notre Dame, IN 46556, USA}

\address{$^2$ Department of Physics, Universidad de Santiago de Chile, Av Victor Jara 3493, Santiago, Chile}
\address{$^3$ Millennium Institute for Research in Optics, Concepci\'on, Chile}
\address{$^4$ Current address: Department of Chemistry, Bowdoin College, ME 04011, USA}

\ead{ycolon.nd.edu; felipe.herrera.u@usach.cl}

\begin{abstract}
    The transmission of strong laser light in nonlinear optical materials can generate output photons sources that carry quantum entanglement in multiple degrees of freedom, making this process a fundamentally
    important tool in optical quantum technology. However, the availability of efficient optical crystals for entangled light generation is severely limited in terms of diversity, thus reducing the
    prospects for the implementation of next-generation protocols in quantum sensing, communication and computing. To overcome this, we developed and implemented a multi-scale first-principles modeling
    technique for the computational discovery of novel nonlinear optical devices based on metal-organic framework (MOF) materials that can efficiently generate entangled light via spontaneous parametric
    down-conversion(SPDC). Using collinear degenerate type-I SPDC as a case study, we computationally screen a database of 114,373 synthesized MOF materials to establish correlations between the structure
    and chemical composition of MOFs with the brightness and coherence properties of entangled photon pairs. We identify a subset of 49 non-centrosymmetric mono-ligand MOF crystals with high chemical and
    optical stability that produce entangled photon pairs with intrinsic $G^{(2)}$ correlation times $\tau_c\sim 10-30$ fs and pair generation rates in the range $10^4-10^{10}$ s$^{-1}$mW$^{-1}$mm$^{-1}$ 
    at 1064 nm. Conditions for optimal type-I phase matching are given for each MOF and relationships between pair brightness, crystal band gap and optical birefringence are discussed. Correlations
    between the optical properties of crystals and their constituent molecular ligands are also given. Our work paves the way for the computational design of MOF-based devices for optical quantum
    technology. 
\end{abstract}
%
\vspace{2pc}
\noindent{\it Keywords}: Metal-Organic Frameworks, Spontaneous Parametric Down Conversion, Entangled Photons, Energy-Time Entanglement
%
%

\section{Introduction}

\begin{figure}[!htbp]
	\centering
    \includegraphics[width=0.65\textwidth]{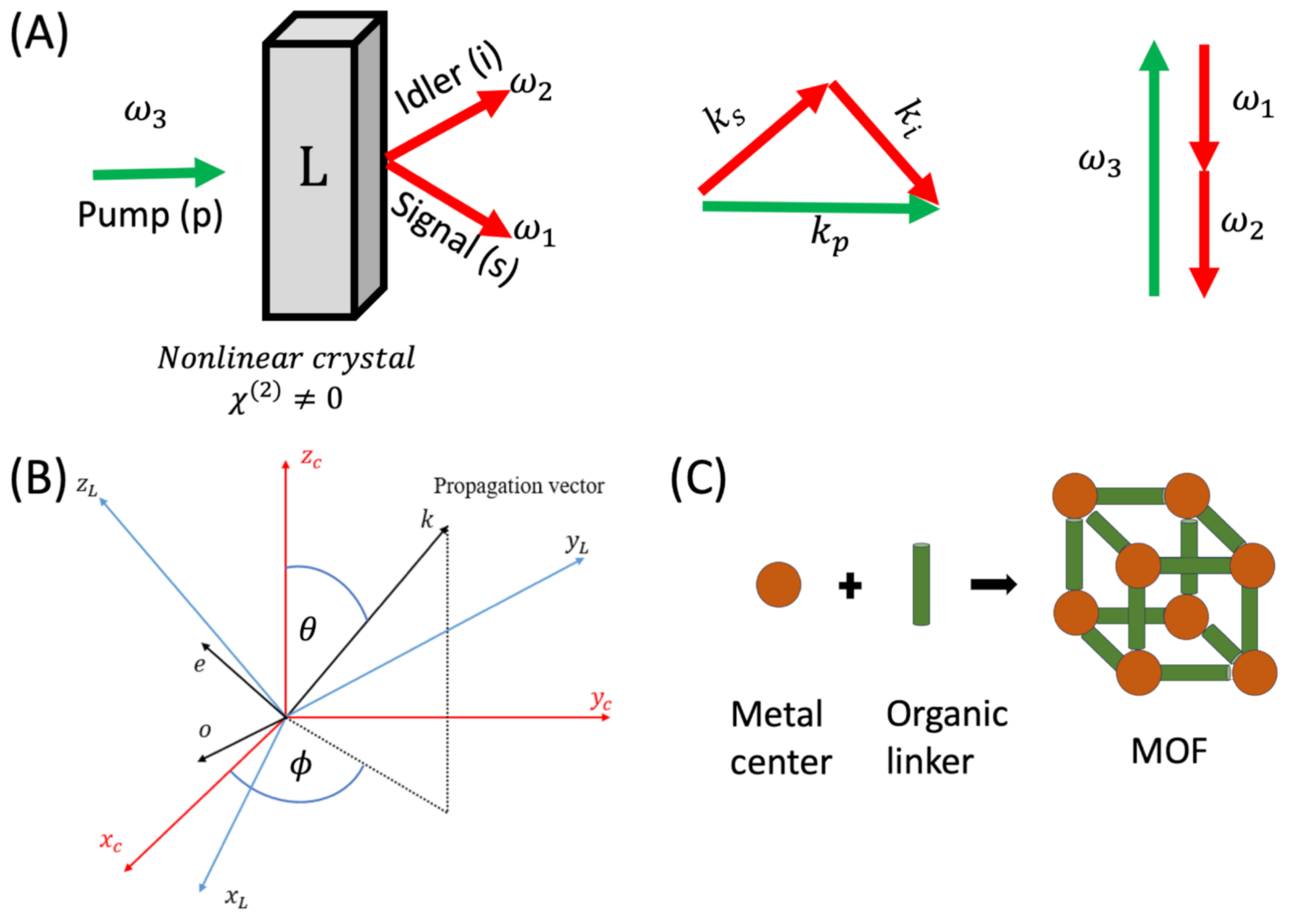}
    \caption{(A) Schematic of SPDC process with momentum and energy conservation. One photon energy $\hbar\omega_3$ splits into two twin photons at energies $\hbar\omega_1$ and $\hbar\omega_2$. 
    (B) Phase matching configuration for type-I SPDC. (C) Sketch of MOF materials.}
    \label{spdc}
\end{figure}

     Entangled photon pair generation is a fundamental technique in quantum optics and quantum information science \cite{pitmann2002,bouwmeester1997,walther2005,tittel2001,irvine2004}. Correlations 
     between entangled photons can be exploited in various applications, such as quantum teleportation \cite{bouwmeester1997}, quantum computing, quantum cryptography \cite{tittel2000,jennewein2000,
     sergienko1999,arrazola2021}, and quantum imaging \cite{pittman1995,angelo2004,str2kalov1995}. Spontaneous parametric down-conversion (SPDC) is a widely used method 
     for generating entangled photon pairs. SPDC occurs when a high intensity laser pump ($p$) interacts with a nonlinear crystal and, as a result of the interaction, pairs of low energy 
     photons are spontaneously generated. The pair components are denoted signal($s$) and idler($i$) \cite{rboyd2008}. The output photons must obey conservation relationship in terms of their angular 
     frequencies and wave vectors ($\omega_p$ = $\omega_s$ + $\omega_i$, $\vec{k_p}$ = $\vec{k_s}$ + $\vec{k_i}$) \cite{rboyd2008,zhang2017}. The SPDC process can be divided into types, degeneracy 
     and direction. In type-I SPDC, the generated photons have the same polarization while in type-II, the output photons have orthogonal polarization. The system is
     collinear if the photons have the same propagation direction as the pump and non-collinear if they propagate in different directions. Figure \ref{spdc} shows a schematic diagram of the SPDC process,
     highlighting energy and momentum conservation. SPDC ($2\omega$ $\rightarrow$ $\omega$) and second harmonic generation ($\omega$ $\rightarrow$ $2\omega$) are conjugate processes in nonlinear optics
     which diﬀer on how they are experimentally implemented \cite{rboyd2008}. For a material to be SPDC efficient and have high performance over a wide range of bandwidths, it should have high nonlinear
     susceptibility, suitable phase matching, group velocity mismatch (GVM) for spectral purity \cite{jin2013}, optical transparency (high band gap), thermal stability and be suitable for practical operation (low 
     cost, availability, ease of use).

    The ability of nonlinear optical materials to be phase matchable depends on whether they belong to non-cubic crystal classes or not. Birefringence phase matching (BPM depends on the difference of refractive indices at
    specific wavelengths and is only possible for anisotropic crystal systems (non-cubic). The non-cubic crystal class is divided into groups: biaxial and uniaxial \cite{rboyd2008}. Triclinic, monoclinic 
    and orthorhombic crystal systems are biaxial, and trigonal, tetragonal and hexagonal are uniaxial. In terms of refractive indices, uniaxial crystals have $n_X$ = $n_Y$ $\ne$ $n_Z$ and biaxial crystals
    have $n_X$ $\ne$ $n_Y$ $\ne$ $n_Z$. For non-linear optical (NLO) applications it is important that the crystal is non-centrosymmetric (NCS) as they have non-zero second order non-linear
    susceptibility
    tensor ($\chi^{(2)}$) which defines the efficiency of the process \cite{rboyd2008}. Uniaxial and biaxial crystals can be further divided into crystal class and the list of all the crystal classes
    which are NCS is known.

    The effective non-linearity, $d_{\rm eff}$, obtained by contracting the $\chi^{(2)}$ with the field polarization of suitable birefringence phase matching condition, is an intrinsic property of 
    materials and plays a crucial role in developing modern optical devices as well as entangled photon sources with high brightness (number of entangled photons) \cite{ruben2021}. The brightness of 
    the signal scales as the square of $d_{\rm eff}$, linearly with the pump power, and linearly with propagation length ($L$) \cite{aramburu2014,kurtz1968} when crystals have a random orientation 
    (powder) and scales as $L^2$ for a single crystals. The effective non-linearity is affected by several material factors such as the electronic structure, chemical composition, optical properties,
    microstructure, temperature and pressure \cite{zuo2022,arrazola2021,wang2020,zhang2022,loang2023,liu2022,li2022,nikogosyan1991,barnes1982,schlarb1993}.

    BBO, LiNbO$_3$, KDP and other leading materials \cite{jin2016,jin2019,jin2023}, have been experimentally realized for nonlinear optical process but have relative disadvantages such as absence of electron delocalization, and the
    inability to adjust their structures \cite{mingabudinova2016}. Organic materials have been proposed to resolve these deficiencies but they lack good mechanical strength and thermal stability. Metal-organic
    frameworks (MOFs) are hybrid materials consisting of metal ions or clusters coordinated to organic ligands to form 1D, 2D or 3D structures, which have shown promise for nonlinear optical process as 
    they can combine the benefits of inorganic and organic compounds \cite{Furukawa2013,Colon2014}. MOFs offer interesting possibilities in terms of synthesis of new structures with high 
    flexibility and tunability, which allows for the optimization of their optical properties. The organic linkers in MOFs can be tuned to achieve a large nonlinear optical response and the metal centers
    can provide high degree of anharmonicity, which can significantly enhance the nonlinear optical response \cite{zuo2022, mingabudinova2016}. MOFs offer high surface area, and are stable under
    various conditions and can be synthesized with a high degree of control over their size, shape and functionality. Several MOF structures have been experimentally synthesized and some of them have
    already shown potential to be used in nonlinear optics \cite{Cleuvenbergen2016-ZIF-8-SHG, Zhang2019-impressive,Chen2020-GIANT,Liu2016a,Chi-Duran2018,Enriquez2019,Garcia2020,duran2022}. A recent
    publication reports the experimental demonstration of three-wave and four-wave mixing processes using millimeter-scale MOF single crystal with well-defined phase matching properties \cite{Hidalgo-Rojas2023}. The
    competitive optical nonlinearities of several polycrystaline MOF samples have stimulated the need to develop a robust computational methodology to screen the MOF databases to understand the structure-property 
    relationship to identify and design potential MOF crystals that are promising.

    In previous work, we developed a multi-scale methodology to compute entangled photon pair properties for uniaxial MOF crystals \cite{ruben2021}. We have also shown that the types of 
    ligand and their arrangement can significantly affect the entangled photon pair properties \cite{sanoj2023}. In this work, we extend our methodology to take into account the calculation of
    $d_{\rm eff}$ for biaxial crystals with appropriate phase matching condition, GVM, spectral acceptance bandwidth, photon correlation ($G^{(2)}$) function and counting rate of entangled 
    photon pairs for the collinear type-I degenerate SPDC. 
    We have carried out high-throughput screening of MOFs with zinc as a metal and one-type of ligand to understand how different structures affect the properties of the entangled photon source. 
    We have selected mono-ligand MOFs with d$^{10}$ metals, specifically Zn, because of its abundance within the MOF structure database. Also, Based on our results, 
    we provide  design rules which can be used to develop new materials which are highly SPDC efficient.


\section{Methods}
\begin{figure}[!htbp]
    \centering
    \includegraphics[width=0.7\textwidth]{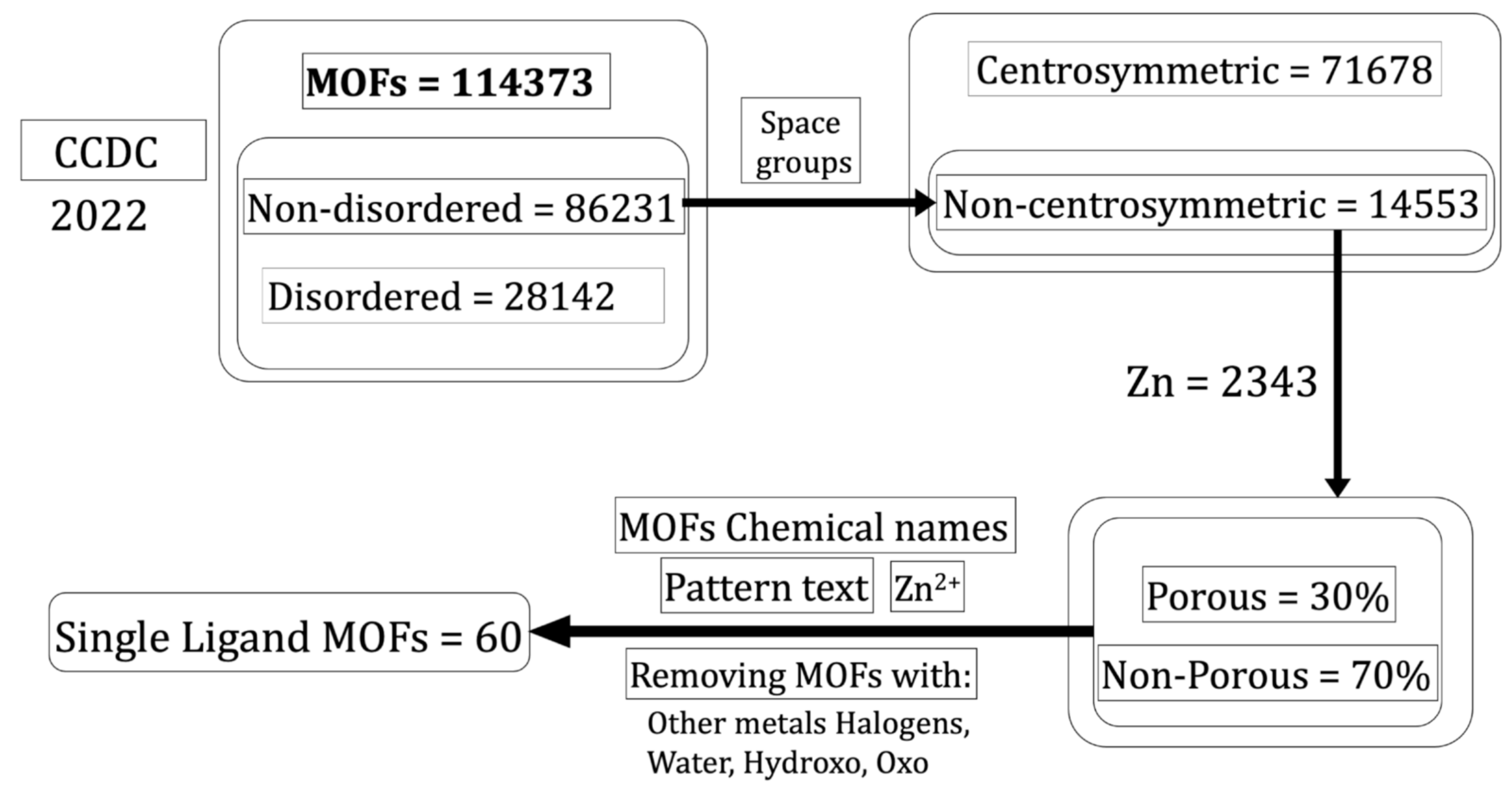}
    \caption{Flowchart of the MOF selection filter. MOFs = 114373 represents total number of MOFs from CCDC database. Zn = 2243 represents total number of non-disordered non-centrosymmetric MOFs with Zinc as metal.}
    \label{scheme}
\end{figure}

    \textbf{MOF selection criteria and computational details:} Figure \ref{scheme} shows the flowchart of the MOF filter. After applying the selection criteria, we obtained a database of mono-ligand 
    MOFs with Zn as a metal. All the MOFs in our database are experimentally synthesized and they come the Cambridge Crystallographic Data Center (CCDC).

    We used VESTA \cite{vesta2011} to manipulate and clean the original cif file, which was obtained from the CCDC. The software suite ISOTROPY \cite{stokes2005} 
    was used to identify the space group of MOF
    crystals. All the crystal structures were optimized using POB-DZVPP basis set with shrink points 3 3 and TOLINTEG 12 12 12 12 18. We also added DFT-D version 4 for the dispersion correction. We 
    used solid-state DFT within the Coupled Perturbed Hartree-Fock/Kohn-Sham method (CPHF/KS) \cite{fmauro2008, fmauro12008} with a PBE functional to compute the dynamical dielectric tensor
    $\epsilon_{ij}(\omega)$ and the dynamical second-order susceptibility tensor $\chi_{ijk}^{(2)}(\omega)$ of the optimized structures. We have used CRYSTAL17 \cite{crystal17} package to carry out all 
    the DFT calculations. We benchmarked these methods in previous work \cite{ruben2021}. We also used available KDP and BBO data to compare as an additional benchmark. The TOLINTEG and SHRINK points 
    sampling were also done to check the energy convergence. Also, 
    the same DFT method was used in previous works for optical measurement of MIRO-101 single crystal \cite{duran2022} and to study the affect of arrangement of ligands and temperature on entangled 
    photon pair properties \cite{sanoj2023}.

    \textbf{Effective nonlinearity estimation}:
    The estimation of the effective non-linearity ($d_{\rm eff}$) of a given crystal is carried out in the orthogonal coordinate system X Y Z in which the dielectric tensor $\epsilon_{ij}$
    is diagonal. A crystal is considered uniaxial (with optic axis Z) if $\epsilon_{XX}$ = $\epsilon_{YY}$ $\ne$ $\epsilon_{ZZ}$, and biaxial (with the optic axis in the plane XZ) if $\epsilon_{XX}
    $ $\ne$ $\epsilon_{YY}$ $\ne$ $\epsilon_{ZZ}$, and one of two relations being fulfilled: $n_X$$>$ $n_Y$ $>$ $n_Z$, or $n_Z$ $>$ $n_Y$ $>$ $n_X$, where $n_X^2$ = $\epsilon_{XX}$, $n_Y^2$ =
    $\epsilon_{YY}$, $n_Z^2$ = $\epsilon _{ZZ}$ are the principal values of the refractive indices. For some biaxial crystals (crystal class 1, 2 and m), the dielectric matrix
    is not diagonalizable \cite{rboyd2008}. In those cases, the trace of dielectric matrix gives the refractive indices.

\begin{figure}[!htbp]
    \centering
    \includegraphics[width=0.75\textwidth]{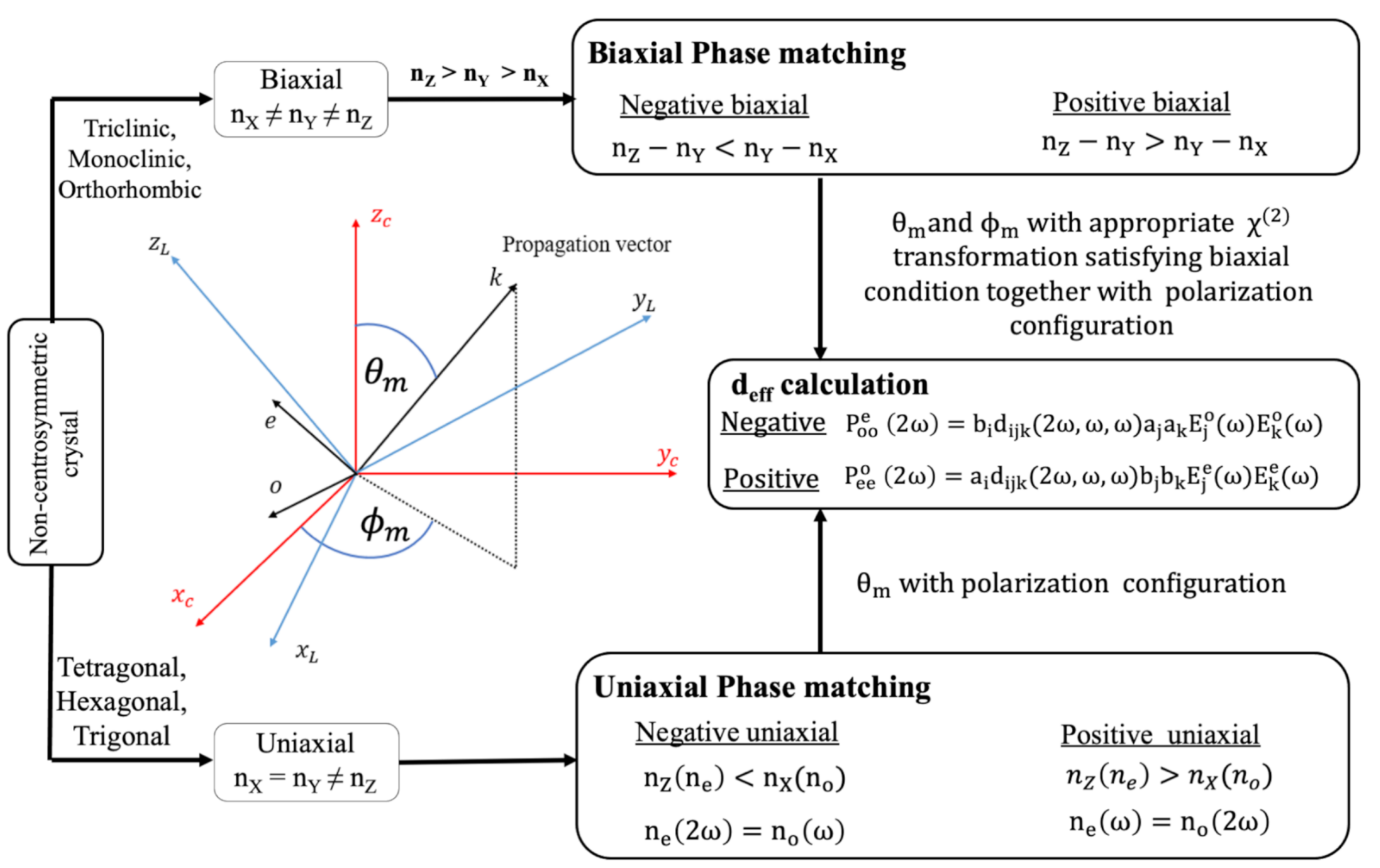}
    \caption{Flowchart for calculating $d_{ \rm eff}$ of any given crystal. $n_X$, $n_Y$ and $n_Z$ are the refractive indices. $n_o$ and $n_e$ are the ordinary and extraordinary refractive indices. The crystal
    optic axis $Z_c$ sets a polar angle $\theta_c$ and azimuthal angle $\phi$ with respect to the propagation direction k. $d_{\rm eff}$ is obtained by contracting the $\chi^{(2)}$ tensor with appropriate
    polarization configuration.}
    \label{method}
    \end{figure}

    Uniaxial and biaxial crystals can be further divided into positive or negative as discussed in the Supporting Information (SI) section 1. For biaxial crystals, it is possible that the principal values of refractive indices in
    the crystallographic
    frame ($a, b, c$) are different than the crystallophysical frame ($X, Y, Z$). Crystallophysical, optical and dielectric frame represent the same concept. Among many widely used crystals such as
    KTP, KNbO$_3$,
    LBO \cite{kato1990}, BiBO \cite{hellwig2000} and $\alpha$-HIO$_3$, the coordinate system $X,Y,Z$ and $a,b,c$ can coincide or misconcide. In KTP, the coordinate systems coincide and we can designate the
    coincidence as $X,Y,Z$ $\rightarrow$ $a,b,c$ whereas for LBO, the assignment $X, Y, Z$ $\rightarrow$ $a,c,b$ is commonly used \cite{kato1990}. It is important to do the necessary transformation of the
    crystallographic frame ($a, b, c$) to satisfy the biaxial condition \cite{kato1990, lin1990} ($n_Z$ $>$ $n_Y$ $>$ $n_X$) and then calculate the $d_{\rm eff}$ for different polarization configurations
    after doing the necessary transformation of the $\chi^{(2)}$ tensor. $d_{\rm eff}$ was estimated using a specific configuration by contracting the full second-order polarizability tensor at the phase 
    matching angles for different crystal structures. Figure \ref{method} shows the flowchart for calculating $d_{\rm eff}$ of any given crystal. Detailed description for calculating birefringence, phase
    matching angles, $d_{\rm eff}$, $G^{(2)}$, and number of entangled photons can be found in SI section 1. Data and code are available on GitHub (\href{https://github.com/snoozynooj/Type-I-SPDC}{Codes 
    and data}) for reproducibility. Although we have explored birefringence phase matching in our study, the MOF crystals used  in our study can be explored for periodic pooling during crystal growth 
    or using other techniques which can achieve quasi-phase matching and can much more efficient. Other types of phase matching can be used for efficient SPDC where materials lack birefringence and are non-linear 
    or domain-disordered \cite{helmy2011}. 

\newpage 

    GVM between signal and pump is important for the spectral purity which is important for heralded single photon sources and entangled photon sources \cite{jin2013}. We 
    have estimate GVM = $1/u_s - 1/u_p$ for all our MOF crystals \cite{shih2003,keller1997}, where $u_s$ is the group velocity of the signal and $u_p$ is the group velocity of pump. More details on 
    calculating GVM can be found in SI section 5.
    
    An important feature for applications of entangled photon pairs is the correlation time between the two generated photons ($\tau_c$). This parameter can be defined as the full width at half maximum 
    (FWHM) of the $G^{(2)}$ correlation function. This correlation function not only depends on the crystal properties, but also on the properties of the experimental set-up such as the bandwidth of 
    the detector \cite{valencia2002,shih2003}. For type-I SPDC the $G^{(2)}$ function can be written as
\begin{eqnarray} 
        G^{(2)}(\tau) \propto \left|\int_{-\infty}^{\infty} d\nu \hspace{0.1cm}     \mathrm{sinc}\left(\frac{\kappa\nu^2 L}{4}\right) e^{-\left(\frac{\nu^2}{\sigma^2}\right)} e^{-i\nu\tau}\right|^2
\label{eq:G2}
\end{eqnarray}
    where $\nu$ correspond to a small detuning frequency from the perfect phase matching, $\kappa$ is the group velocity dispersion (GVD), $L$ is the crystal length and $\sigma$ is the bandwidth of 
    the detector. We have used single crystal length of 1 mm for our calculations. We chose to normalize the $G^{(2)}$ correlation function because the correlation time of the two generated photons does not depend on 
    the amplitude of the function; instead, it relies solely on the full-width at half maximum (FWHM) of the function. In that way we can directly compare plots to determine which crystal exhibits a 
    wider correlation function, indicating a longer correlation time. 
 
    We also assume a Gaussian model for the transverse field profile of the pump and the two generated photons (signal and idler), collecting the down-converted light into single-mode fibers such that 
    the transverse width of the two
    generated  photons is the same $\sigma_1 = \sigma_2$. The single-mode counting rate for photon pairs generated by type-I SPDC is given by \cite{schneeloch2019introduction}
 
\begin{eqnarray} 
     R &=&  \frac{|E_{p}^{0}|^{2}(d_{\rm eff})^2L^{2}}{2\pi c^{2}} \frac{n_{g1}n_{g2}}{n_{1}n_{2}}\left|\frac{\sigma_{p}^{2}}{\sigma_{1}^{2}+2\sigma_{p}^{2}}\right|^{2} \int d\omega_{1}\omega_{1}(\omega_p - \omega_1) \mathrm{sinc}^{2} \left(\frac{\Delta k L}{2}\right)
    \label{eq:counting_rate}
\end{eqnarray}
    where $\sigma_p$ and $\sigma_1$ are the transverse width of the Gaussian profiles associated with the pump and signal or idler photon respectively. $n_1$ and $n_2$ are the signal and idler refractive indices,
    $n_{g1}$ and $n_{g2}$ correspond to the group indices of the signal and idler frequency. $c$ is the velocity of the light, $|E^0_p| =\frac{|D^0_p|}{e_o n^2}$  is proportional to the monochromatic 
    pump beam peak magnitude $|D_p^0|$, and $P = c |D_p^0|^2 \pi \sigma_p^2 /{n^3 \epsilon_0}$ is the amount of power delivered by a Gaussian pump beam. The phase mismatch $\Delta k$ is 
    given by
 
\begin{equation}
         \Delta k \approx \kappa \left( \omega_1 - \omega_p   \right)^2,
\end{equation}
    where $\omega_1$ correspond to the frequency of the signal and idler photon and $\omega_p$ is the fixed pump frequency.

    We have used the pump wavelength of 532 nm for our study and MOFs whose band gap is greater than $\sim$2.33 eV. This is to ensure that there is no absorbance at 532 and 1064 nm. After applying the selection 
    criteria on the band gap, we found 49 MOFs that are suitable for this study with exception of ECIWAO whose band gap is $\sim$2.28 eV.

\section{Results and discussions}
\begin{figure}[!htbp]
    \centering
    \includegraphics[width=0.70\textwidth]{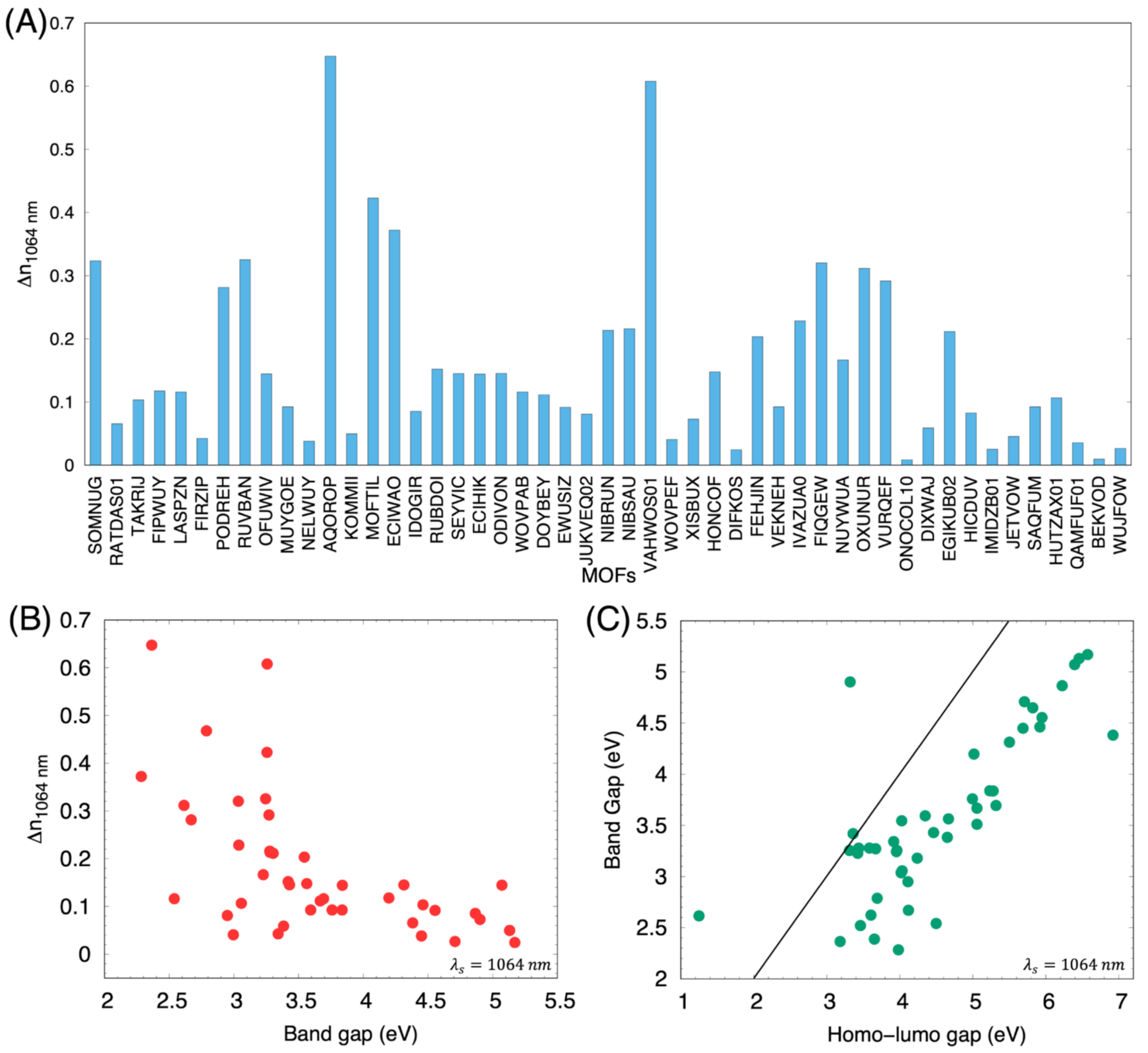}
    \caption{(A) Birefringence $\Delta n$ for different MOF crystals. (B) $\Delta n$ as a function of band gap for different MOF crystal. (C) Correlation plot between between the homo-lumo gap of the
    ligands and band gap the crystal which has this ligands. The black line represents the perfect correlation between the band gap and HOMO-LUMO gap.}
    \label{Figure1}
\end{figure}

    Birefringence ($\Delta n$ = $n_x$ - $n_z$) plays a crucial role in achieving phase-matching conditions for nonlinear materials. The variations in $\Delta n_{1064 nm}$ for all the crystals used in our 
    study are depicted in Fig. \ref{Figure1}A. The MOF crystal AQOROP has the highest $\Delta n$ and ONOCOL01 has the smallest value. Details about crystal class, band gap and birefringence are shown in SI section 2.

    Figure \ref{Figure1}B illustrates the relationship between $\Delta n$ and the band gap for different crystals. We find that $\Delta n$ decreases with increased band gap. Furthermore, Fig.
    \ref{Figure1}C displays the correlation between the HOMO-LUMO gap of the ligand monomers and the band gap the crystals that contain these ligands. A strong correlation is evident, indicating 
    that HOMO-LUMO gap influences the  band gap of the crystal, including a red shift that results from interactions within the crystal. This
    observation serves as a valuable design criterion for crystals and provides the ability to control the band gap. The ability 
    to control the band gap is very important, particularly in getting  transparent crystals. Moreover, $\Delta n$ holds a key role in achieving phase matching over a wider bandwidth. These features are
    critical for the advancement of efficient nonlinear optical materials. The careful selection of ligands emerges as a powerful tool to fine-tune these optical properties for desired specifications. 

    The observations discussed above can be explained by considering the polarizability ($\alpha$). Polarizability refers to the ease with which the electron cloud of a material can be distorted by an
    external electric field and is directly linked to the refractive index. A material with higher polarizability, indicating a greater ability for charge separation, will have a higher refractive index 
    and higher $\Delta n$ \cite{weinbing2022}. Charge separation and delocalization of electrons are closely related to each other. Highly delocalized electrons lead to higher polarizability 
    which affects the refractive index and $\Delta n$. Furthermore, electron delocalization affects the HOMO-LUMO gap for ligand molecules and the crystal band gap. Higher delocalization of electrons,
    gives smaller band gaps \cite{yan2013}. Therefore, the energy gap and birefringence are inversely related to each other.

    Birefringence and phase-matching are related with each other. An increase in birefringence increases the phase-matching  wavelength range which can significantly affect the nonlinear efficiency. It 
    should be noted that if birefringence is too large, serious walk-off effect can occur and nonlinear efficiency can significantly decrease even if the nonlinear coefficients are higher \cite{mori2002,
    zhang2017}. 
    If the birefringence is too small, phase-matching will not occur. We have calculated the phase-matching angles of all the crystals depending on whether a crystal is uniaxial or biaxial 
    at 1064 nm (see Method section). We benchmarked our method by reproducing phase-matching angles calculations for KTP \cite{yao1984} and BiBO \cite{hellwig2000,huo2015aa} as shown in SI section 3.

\begin{figure}[!htbp]
    \centering
    \includegraphics[width=0.7\textwidth]{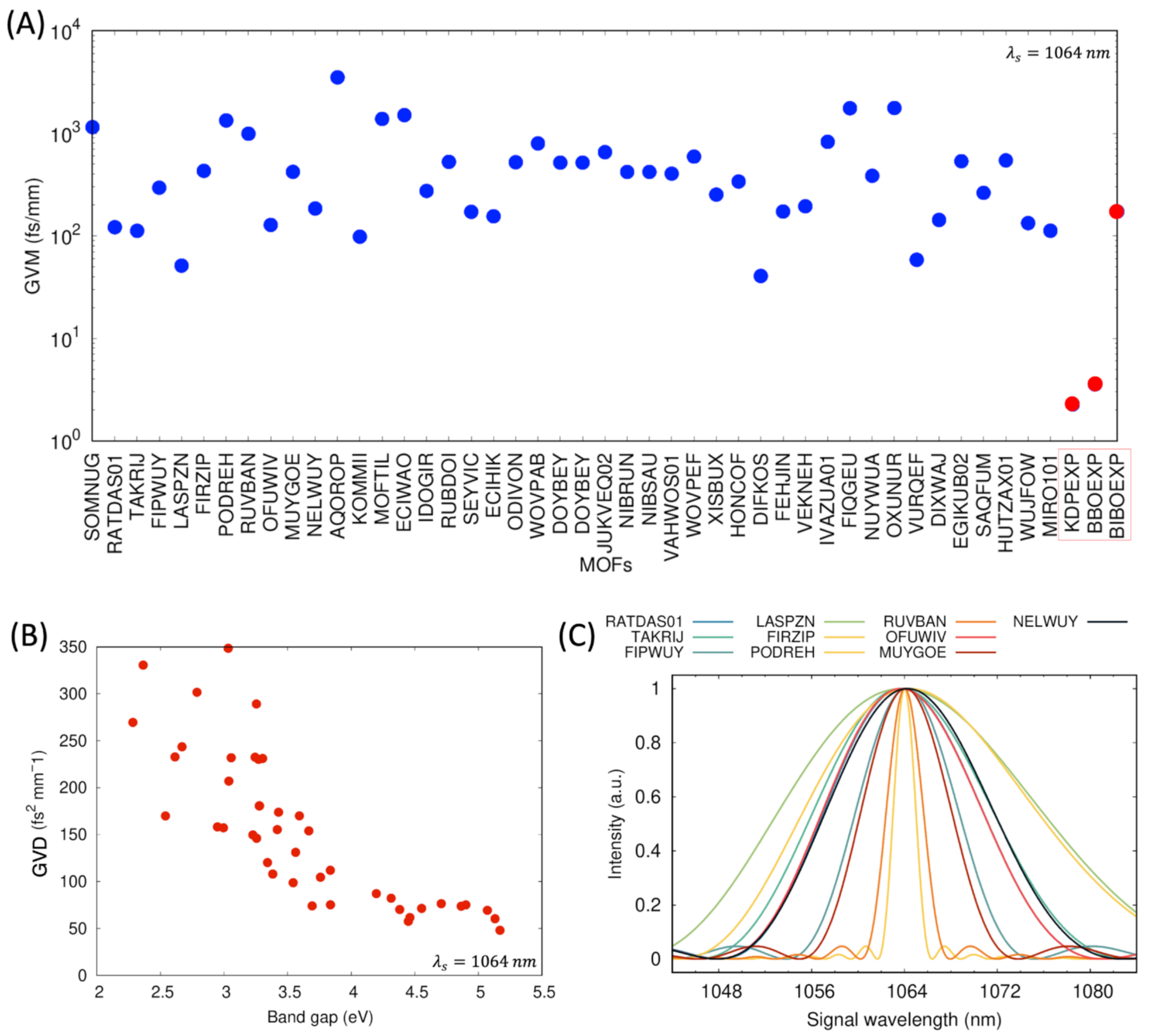}
    \caption{(A) GVM of different crystal at pump wavelength 532 nm. Blue circles represent the GVM of all the MOFs used in our study using calculated refractive indices and red circles represent the GVM  
    of KDP, BBO and BiBO using experimental refractive indices. (B) GVD as a function of band gap for different MOF crystals. (C) spectral acceptance bandwidth plot for crystals belonging to crystal class 2.}
    \label{Figure10}
\end{figure}

    \textbf{Dispersive properties of nonlinear crystals}: 
    Nonlinear crystals dispersive properties are of great importance in the femtosecond pumping regime. This is due to their effects on the pulse width, the effective interaction length of the pump pulses, 
    and the conversion efficiency. Temporal walk-off given by GVM and temporal broadening given by GVD impose limits on the applicability of crystal length \cite{Akbari2013}. We have estimated the GVM \cite{keller1997,ghotbi2004} and 
    Group velocity dispersion (GVD) for the crystals used in our study. Figure \ref{Figure10}A shows the GVM of all the crystals at a signal wavelength of 1064 nm. We have also included GVM of KDP, BBO, 
    and BiBO using the experimental data \cite{Umemura2007,Eimerl1987,Zernike1964} and are shown in Fig. \ref{Figure10}A as KDPEXP, BBOEXP, and BiBOEXP. Figure \ref{Figure10}B shows the GVD of all the MOFs 
    as a function of the band gap. We observe that the GVD decreases with an increasing band gap. Figure \ref{Figure10}C shows the spectral acceptance bandwidth of the phenomenon at a signal wavelength 
    ($\lambda_s$ = 1064 nm) using the joint spectral function for crystals belonging to crystal class 2. Details about calculating GVM, GVD, and spectral acceptance bandwidth are discussed in SI section 5. 
    We have also included the spectral acceptance bandwidth plots of different crystal classes and GVM as a function of signal wavelength using the experimental refractive indices data of KDP, BBO, and BiBO, 
    including the calculated BiBO GVM, in SI section 5. 

\begin{figure}[!htbp]
    \centering
    \includegraphics[width=0.93\textwidth]{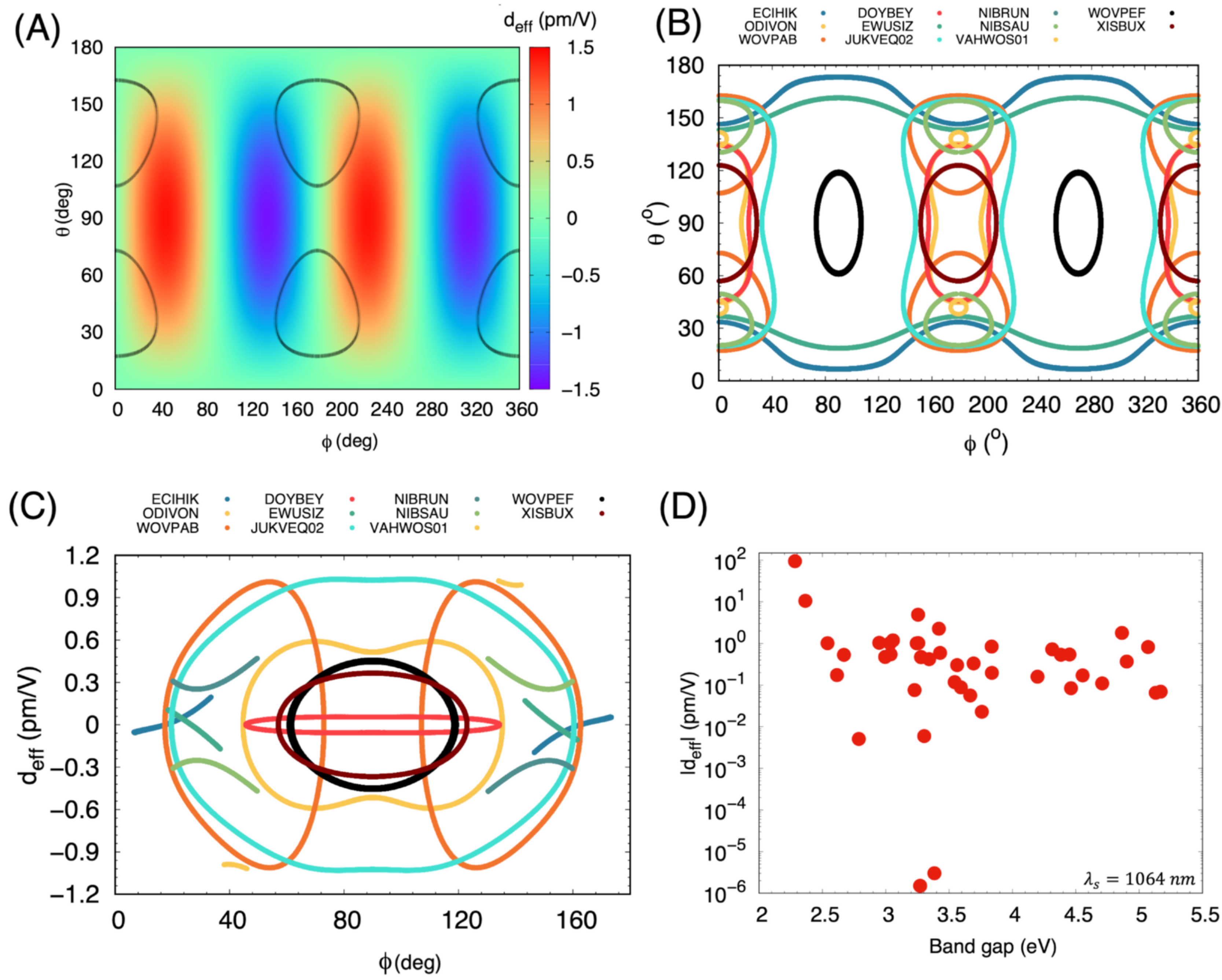}
    \caption{(A) 2d map of $d_{\rm eff}$ as function of $\theta$, $\phi$ of WOVPAB MOF crystal belonging to 222 crystal class. Black lines shows all the possible phase matching angles ($\theta_m$ and $\phi_m$
    ). (B) Phase matching angles ($\theta_m$ and $\phi_m$) of biaxial crystals belonging to crystal class 222. (C) $d_{\rm eff}$ of crystals belonging to crystal class 222 as function of phase matching angle,
    $\theta_m$. $\phi_m$ is not shown for clarity. (D) $d_{ \rm eff}$ as function of band gap for different crystals. Crystals whose $d_{\rm eff}$ values are closer to $10^{-6}$ pm/V belong to the 422 
    crystal class.}
    \label{Figure61}
\end{figure}

    For NLO materials to be technologically useful, it is not enough for a material to be highly efficient at a given wavelength but also should be phase matchable. The 
    phase matching angle together with the $\chi^{(2)}$ tensor determine the $d_{\rm eff}$ of the crystal. Figure \ref{Figure61}A shows the overlay of possible phase matching angles, shown with black lines, 
    with the 2D map of $d_{\rm eff}$ for all the $\theta$ and $\phi$ for a WOVPAB crystal belonging to crystal class 222. It is important to note that although some combination of $\theta$ and $\phi$ 
    can give high $d_{\rm eff}$, those angles may not be accessible due to phase-matching conditions. Figure \ref{Figure61}B shows the phase matching angles ($\theta_m$ and $\phi_m$) of biaxial 
    crystals belonging to crystal class 222. We observe that although the crystals belong to the same crystal class, they have very different phase-matching angular profiles.

    The effective non-linearity is an intrinsic property of a materials and plays a crucial role in developing modern optical 
    devices as well as entangled photon sources with high brightness (number of entangled photons). The source brightness scales with $d_{\rm eff}^2$, and $d_{\rm eff}$ is used to parameterize the 
    effective Hamiltonian for the SPDC process. In order to calculate $d_{\rm eff}$ at 1064 nm for different crystals, we have divided MOFs into uniaxial and biaxial classes. We have used the phase
    matching condition to find phase matching angles, $\theta_m$ for uniaxial, and $\theta_m$ and $\phi_m$ for biaxial at 1064 nm, and used the appropriate polarization configuration to find the $d_
    {\rm eff}$ (see Method Section and SI Section 2 for transformation between crystallophysical and crystallographic frame to satisfy biaxial condition). The calculated $\chi^{(2)}$ for all the MOFs 
    used in our study are given in SI Section 4. Figure \ref{Figure61}C shows $d_{\rm eff}$ as function of $\theta$ ($\phi_m$ is not shown for clarity) for MOF crystals belonging to crystal class 
    222. The sign of $d_{\rm eff}$ is unimportant and it is the profile of $d_{\rm eff}^2$ that dictates the conversion efficiency along a given phase matching direction. Phase matching angles ($\theta_m$) of 
    all the uniaxial crystals are shown in SI Section 2 and phase matching angles ($\theta_m$ and $\phi_m$) of all the biaxial crystals used in our study are shown in the SI Section 3.

    Figure \ref{Figure61}D shows the absolute maximum value of $d_{\rm eff}$ of all the crystal as a function of their band gaps. We observe that $d_{\rm eff}$ decreases with increasing band gap.
     This is because materials with smaller band gaps enable greater degree of overlap between the wavefunctions of electrons and holes and leads to a higher probability on nonlinear optical
    interactions which can significantly affect the efficiency of the non-linear process. Materials with smaller band gaps lead to higher nonlinearities as compared to higher band gap because their
    electronic structures enable a greater degree of optical mixing and frequency conversion. We also observe that MOF crystals belonging to crystal class $422$ (uniaxial crystal) have $d_{\rm eff}$ 
    = 0 pm/V (see Fig. \ref{Figure61}D) although it has small $\chi^{(2)}$ as shown in Fig. S2. This agrees well with the literature as crystals belonging to the 422 crystal class will have vanishing $d_{\rm eff}$ if Kleinman's
    symmetry is valid \cite{zernike1973, dmitriev1999}. 

\begin{figure}[!htbp]
    \includegraphics[width=1.0\textwidth]{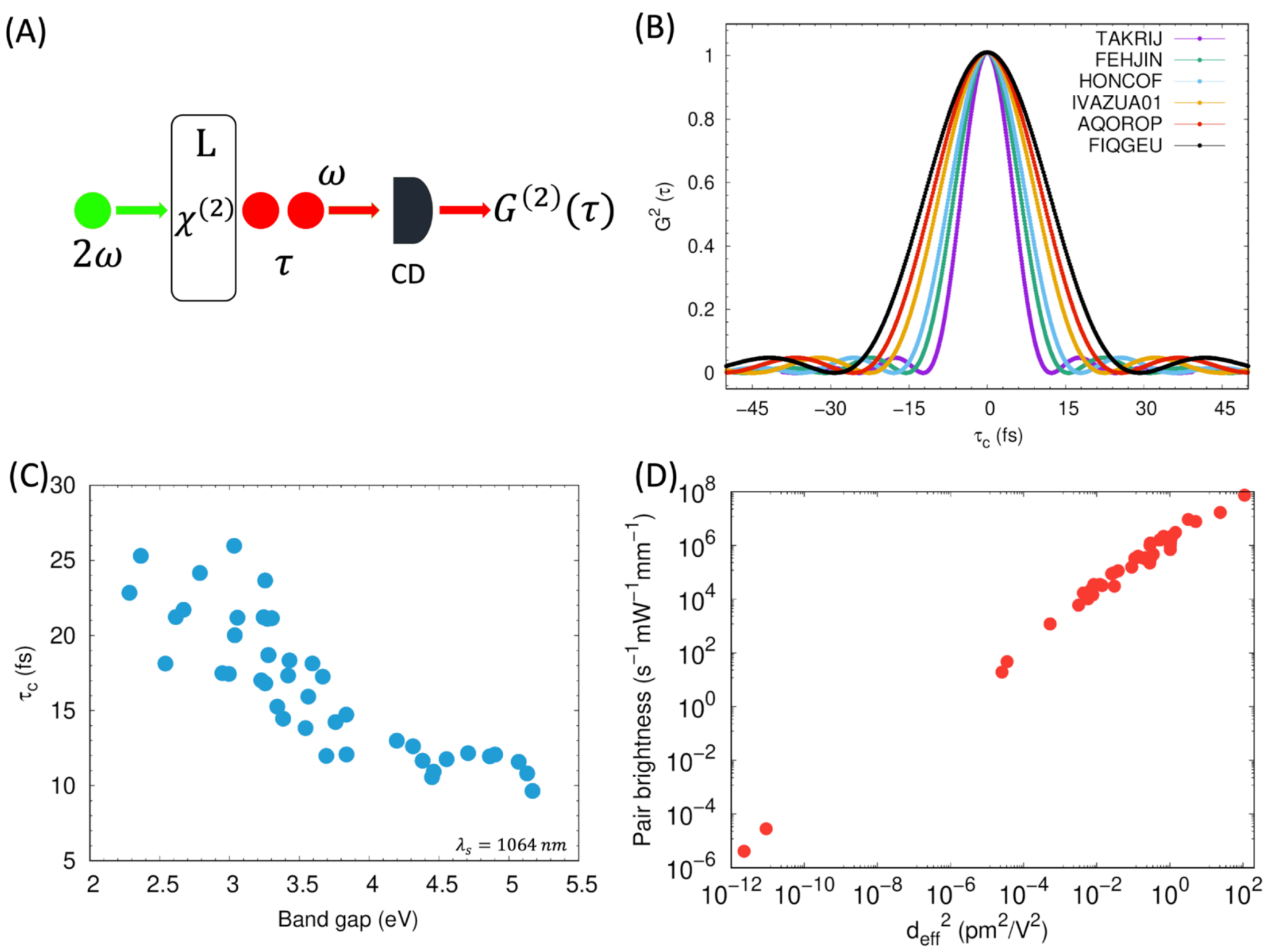}
    \caption{(A) Visible photons (green) are pumped into a NCS MOF single crystal with propagation length L. Entangled photon pairs at half the pump frequency (red), produce from the crystal with a small
    delay $\tau$. Photon pairs are collected with a coincidence detector (CD), and the delay times are post-processed to give a two-time correlation function $G^{(2)}(\tau)$. A special case of type-I SPDC 
    where the signal and idler are coming to the same detector and the difference between the arrival time of the photons represents the correlation time. (B) $G^{(2)}(\tau)$ of 
    selected MOFs. (C) Entangled photon correlation time as a function of band gap for all the crystals. (D) Pair brightness as a function of $d_{\rm eff}^2$.}
    \label{figure7}
\end{figure}

\begin{figure}[!htbp]
     \centering
     \includegraphics[width=1.0\textwidth]{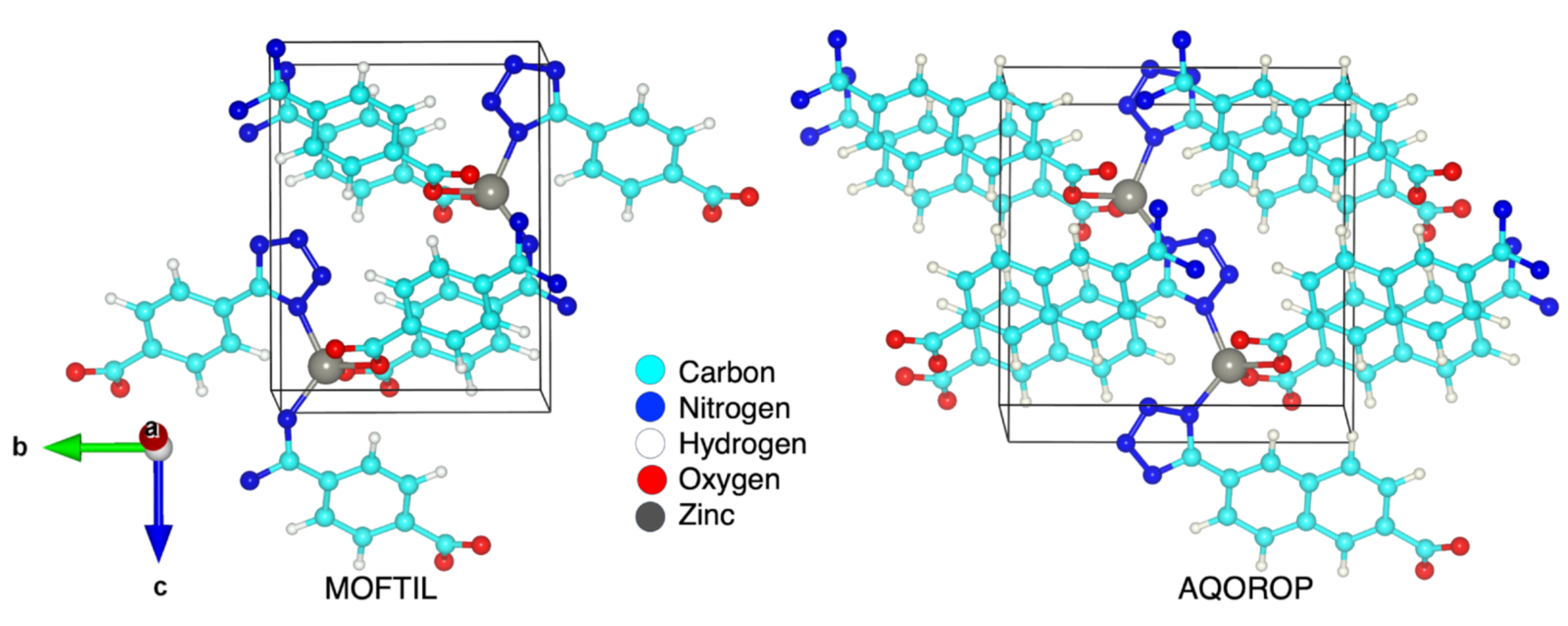}
     \caption{Optimized MOFTIL and AQOROP structures}
     \label{Figure11}
 \end{figure}

     \textbf{Entangled photon pair properties:}
     The $G^{(2)}$($t_1-t_2$) intensity correlation function gives the probability of detecting two-photon pairs at different time $t_1$ and $t_2$ (see Method section for more details). Figure
     \ref{figure7}A shows the schematic of coincidence setup for measuring the two-time correlation function $G^{(2)}(\tau \equiv t_1-t_2)$. Figure \ref{figure7}B shows $G^{(2)}(\tau)$ for some of the 
     MOFs used 
     in this study (see SI Section 4 all the MOFs $G^{(2)}$). The FWHM of $G^{(2)}(\tau)$ gives the pair correlation time ($\tau_c$). Figure \ref{figure7}C shows $\tau_c$ as function of band gap for 
     all the crystals. We observe that $\tau_c$ decreases with increasing band gap. This is due to higher and smaller degrees of overlap between the wavefunctions of electrons and holes for
     smaller and larger band gaps. The degree of correlation between the generated photons is relevant in applications such as quantum computing, quantum key distribution, quantum teleportation, and
     quantum cryptography \cite{tittel2000,jennewein2000,arrazola2021,bouwmeester1997}. The calculated correlation time for all the MOF crystals are in femtosecond range which is extremely challenging 
     to measure them using photodetectors as they have resolution in picosecond regime. The fiber optics play a key role in making correlation time larger to nanoseconds from femtoseconds 
     \cite{G2Fiber,valencia2002}. 
     We find that AQOROP ($\sim 27.55$ fs), ECIWAO ($\sim 24.87$ fs) and MOFTIL ($\sim 25.76$ fs) have the largest autocorrelation times.
 \newpage
     The number of entangled photon pairs generated in SPDC is an important parameter for entanglement-based applications \cite{shih2003}. Figure
     \ref{figure7}D shows the number of entangled photon pairs generated by all the crystals as function of their $d_{\rm eff}^2$. We found that AQOROP ($\sim 7.47 \times 10^7$ s$^{-1}$ mW$^{-1}$ mm$^{-1}$),
     ECIWAO ($\sim 6.99 \times 10^9$ s$^{-1}$ mW$^{-1}$ mm$^{-1}$) and MOFTIL ($\sim 1.69 \times 10^7$ s$^{-1}$ 
     mW$^{-1}$ mm$^{-1}$) have the largest photon conversion rates. Figure \ref{Figure11} shows the components of AQOROP and MOFTIL MOF crystals.
\newpage
     All the results discussed above are with signal wavelength of 1064 nm. From all the crystals we have used in our study, we could not estimate $d_{\rm eff}$ for QAMFUF01 and BEKVOD crystals although 
     they have high $\chi^{(2)}$ as shown in Figs. S3 and S4, because we were not able to find any phase matching angle at 1064 nm. 
\begin{figure}[!htbp]
    \includegraphics[width=1.0\textwidth]{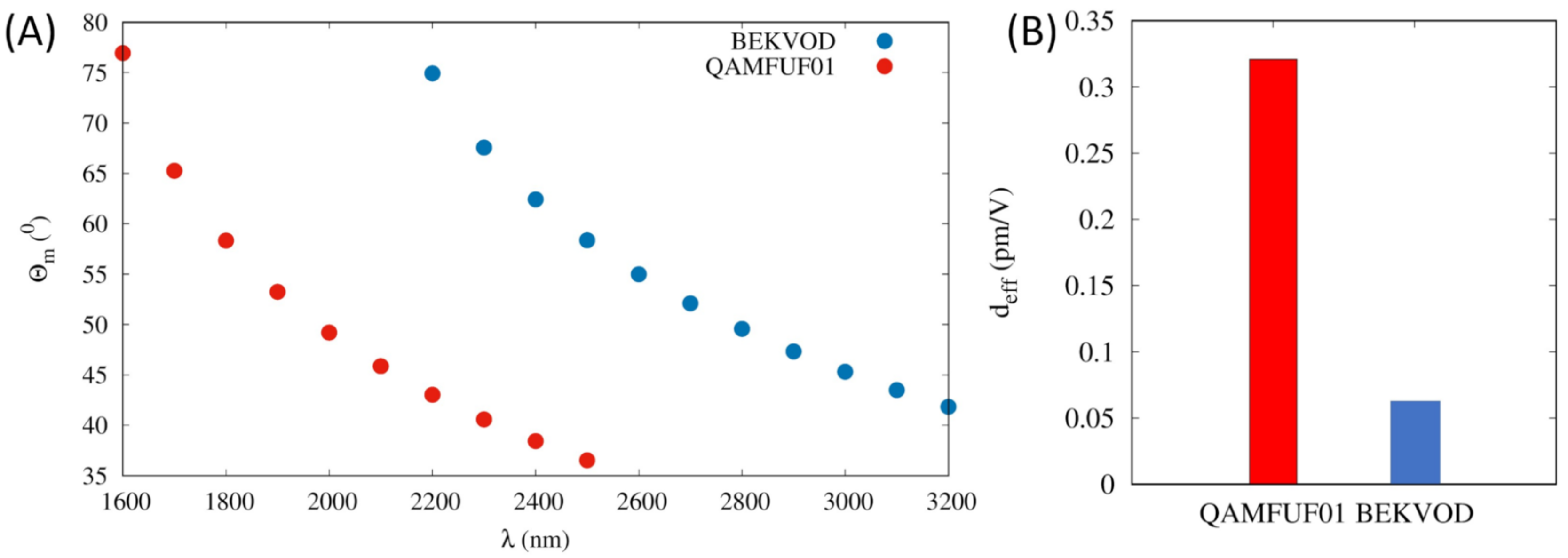}
    \caption{(A) Phase matching angle ($\theta_m$) for QAMFUF01 and BEKVOD crystals for a range of signal wavelengths. (B) Effective nonlinearity ($d_{\rm eff}$) for for QAMFUF01 crystal at 1600 nm and 
    BEKVOD crystal at 2200 nm.}
    \label{Figure5}
\end{figure}
      In order for these crystals to be phase-matchable, we plotted the phase-matching angle as a function of wavelength. Figure \ref{Figure5}A shows how $\theta_m$ changes with 
      signal wavelength for QAMFUF01 (1600 - 2500 nm) and BEKVOD (2200 - 3200 nm). We then calculated $\chi^{(2)}$ for QAMFUF01 and BEKVOD crystals at 1600 and 2200 nm. Figure 
      \ref{Figure5}B shows the $d_{\rm eff}$ of these two crystals at 1600 and 2200 nm.
\newpage
      Having nonlinear optical materials to be operational over wider bandwidth is an important requirement for photonics application. We have shown in Table \ref{table1} how the photon conversion rate 
      is affected as a function of pump wavelength. We chose MOFTIL because it is the best crystal for entangled photon sources with pair conversion rate of $1.7 \times 10^7$ s$^{-1}$ mW$^{
      -1}$ mm$^{-1}$, and birefringence of 0.42 at 1064 nm and band gap of 3.2 eV. We observe that $d_{\rm eff}$ and entangled  photon pairs decrease with increasing pump wavelength (decreasing photon
      energy).

\begin{longtable}{|p{2.0cm}|p{2cm}|p{1.5cm}|p{3.0cm}|p{1.5cm}|p{2.0cm}|}
\caption{$d_{\rm eff}$, number of entangled photon pairs and photon pair correlation time at different signal wavelength for MOFTIL crystal.}
\label{table1}\\
\hline
    MOF &  $\lambda_s$ (nm) ($\lambda_p$=$\lambda_s$/2)  & $d_{\rm eff}$  (pm/V) & R $ $(s$^{-1}$mW$^{-1}$mm$^{-1}$) & $\tau_c$ (fs) & GVM (fs/mm) \\
\hline
    MOFTIL   & 909  & 6.24 & 31.36 $\times$ $10^6$ & 26.63 & 2615.67  \\
\hline
    & 1064 & 4.88 & 16.13 $\times$ $10^6$  & 23.66 & 1410.07 \\
\hline
    & 1100 & 4.70 & 14.32 $\times$ $10^6$ & 23.11 &  1269.64 \\
\hline
    & 1546 & 3.61 & 5.42 $\times$ $10^6$ & 18.61 & 480.95 \\
\hline
\end{longtable}

\begin{figure}[!htbp]
	\centering
    \includegraphics[width=0.6\textwidth]{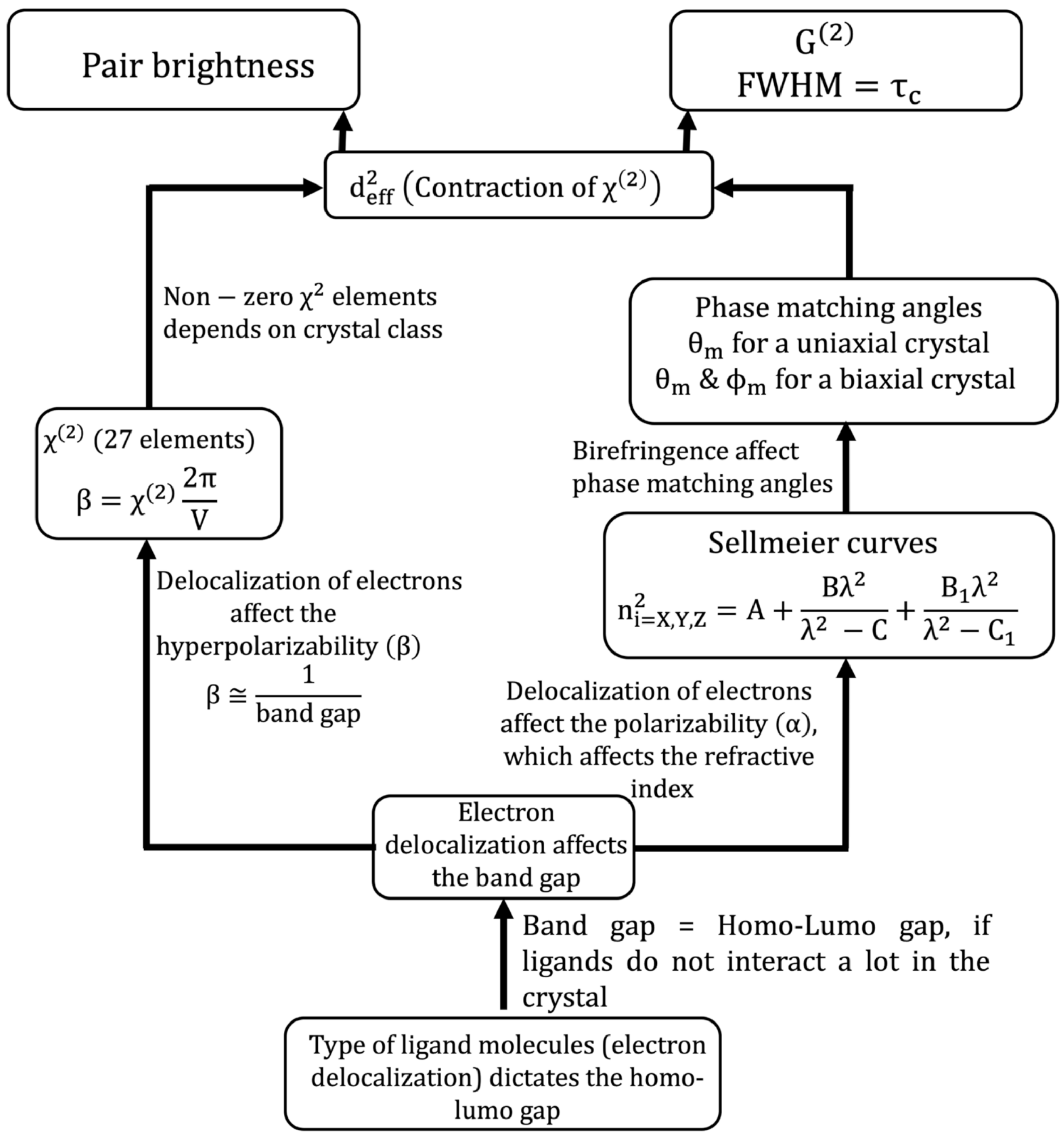}
    \caption{Flowchart showing how different quantities are correlated with each other.}
    \label{figure8}
\end{figure}

\newpage
    In summary, the band gap, birefringence, signal wavelength, phase matching angles, effective non-linearity (contracted form of $\chi^{(2)}$ tensor), number of entangled photons are correlated with each
    other. Figure \ref{figure8} shows how all of these quantities are interrelated and how these correlation can be exploited for designing new materials which will be SPDC efficient. 

\newpage
    \section{Conclusion}
    We have successfully explored MOF database for mono-ligand MOFs with Zn as a metal center for generating entangled light via collinear degenerate Type-I SPDC. All the MOFs in our database is 
    experimentally synthesized and come from CCDC after applying MOF filter as discussed in the 
    method section. We have systematically developed and
    implemented a multiscale computational technique to the selected MOFs (uniaxial or biaxial) to identify the features that are important for high brightness and large photon correlation times.
    We found MOFTIL crystal to be the best candidate entangled photon source with photon conversion rates of $1.7 \times 10^7$ pairs s$^{-1}$ mW$^{-1}$ mm$^{-1}$, and birefringence of 0.42 at 1064 nm 
    and band gap 
    of 3.2 eV. We observe that the band gap of the crystals is related to the properties such as birefringence (an important parameter for angle phase matching), effective nonlinearity and the entangled
    photon pair properties. Band gap can be related to the HOMO-LUMO gap of the ligand, assuming that the ligands do not interact significantly in the crystal, which can be a design criteria for
    preparing new MOF crystals with desired band gaps. Advanced entangled photon sources are crucial for photonic quantum technology and MOF nonlinear materials show great promise for modern 
    nonlinear applications.

    \section*{Acknowledgments}
    YJC thanks the University of Notre Dame for financial support through start-up funds. SP, RAF and FH are supported by ANID through ANID Fondecyt Postdoctoral 3220857, Fondecyt Regular 1221420 and 
    Millennium Science Initiative Program ICN170\_12. We thank center for research computing (CRC) at University of Notre Dame (UND) for computing resources.

    \section{Data and Codes Availability}
    All the data for Sellmeier curves and second-order susceptibility tensor obtained with CRYSTAL17 are available on GitHub. Codes to calculate phase-matching angles, effective nonlinearity,
    correlation function and number of entangled photon pairs for Type-I SPDC are available on GitHub. Link to GitHub page is \href{https://github.com/snoozynooj/Type-I-SPDC}{GitHub-Data-Codes}.

    \section{References}

    \bibliographystyle{iopart-num-mod}
    \bibliography{screening.bib}

\section{Supplementary Information}


\section*{Supplementary material}

# First-Principles Screening of Metal-Organic Frameworks for Entangled Photon Pair Generation

**Sanoj Raj[1,4], Simón Paiva[2], Rubén A. Fritz[2], Felipe Herrera[2,3] and Yamil J. Colón[1]**

[1] Department of Chemical and Biomolecular Engineering, University of Notre Dame, IN 46556, USA

[2] Department of Physics, Universidad de Santiago de Chile, Av Victor Jara 3493, Santiago, Chile

[3] Millennium Institute for Research in Optics, Concepción, Chile

[4] Current address: Department of Chemistry, Bowdoin College, ME 04011, USA

E-mail: ycolon.nd.edu; felipe.herrera.u@usach.cl

## Contents

## 1. Method

**Effective non-linearity:**

The electric field component in the ordinary axis is given by $E_j^o(\omega) = (a_j)E^o(\omega)$ and the electric field component in the extraordinary axis is given by $E_j^e(\omega) = (b_j)E^e(\omega)$, where,

$$(a_j) = \begin{pmatrix} sin\phi \\ -cos\phi \\ 0 \end{pmatrix}, (b_j) = \begin{pmatrix} -cos\theta cos\phi \\ -cos\theta sin\phi \\ sin\theta \end{pmatrix}. \quad (1)$$

<u>Uniaxial crystal</u>: For type-I negative uniaxial, $d_{\text{eff}}$ is calculated as

$$P_{oo}^e(\omega_3) = b_i d_{ijk}(\omega_3, \omega_2, \omega_1) a_j a_k E_j(\omega_2) E_k(\omega_1), \quad (2)$$

and for a positive uniaxial crystal, we use

$$P_{ee}^o(\omega_3) = a_i d_{ijk}(\omega_3, \omega_2, \omega_1) b_j b_k E_j(\omega_2) E_k(\omega_1), \qquad (3)$$

$P(\omega_3)$ is the dielectric polarization of the crystal at the pump frequency $\omega_3 = \omega_1 + \omega_2$ ($\omega_3 = 2\omega$ and $\omega_2 = \omega_1 = \omega$). A conventional *e-oo type*-I condition was used [1] for negative uniaxial crystal, where the entangled signal and idler are polarized along the ordinary axis ($o$) and the pump is polarized along the extraordinary axis ($e$). For a positive uniaxial crystal, a conventional *o-ee type*-I condition was used

<u>Biaxial crystals</u>: To estimate the $d_{\text{eff}}$ for biaxial crystals, it is important that the chosen crystallographic axes a, b, c does not differ from the crystallophysical (optical) one (X, Y, Z), for which one of two relations ($n_X > n_Y > n_Z$, or $n_Z > n_Y > n_X$) is necessarily fulfilled [2, 3, 4, 5]. For biaxial crystal, the choice of satisfying relations $n_X > n_Y > n_Z$, or $n_Z > n_Y > n_X$) is actually superfluous [4, 2, 6, 7]. We have satisfied the relation $n_Z > n_Y > n_X$ and necessary transformation of tensor element is done to calculate the $d_{\text{eff}}$ using Eqns. 2 and 3 for negative and positive biaxial crystal. The phase matching direction which gives the maximum $d_{\text{eff}}$ among all possible phase matching directions, solving equation 5, is termed the optimum phase matching direction. We use the maximum value of $d_{\text{eff}}$ for other analysis.

It is important to note that in order to calculate $d_{\text{eff}}$, we have to know whether the crystal is uniaxial (positive/negative) or biaxial (positive/negative) and what are the phase matching angles ($\theta_m$ for uniaxial, $\theta_m$ and $\phi_m$ for biaxial). The difference in the phase matching angle is due to one axis enabling symmetry of revolution in uniaxial crystals compared to biaxial crystals which lack that symmetry. All the discussion about a crystal being positive/negative uniaxial or positive/negative biaxial and their phase matching angle calculations are discussed below. The angles defined are according to the positive nonlinear optics frame convention of Roberts [7].

**Positive/negative crystal and Birefringence**: The value of birefringence in uniaxial crystal is the difference between $n_o(n_X)$ and $n_e(n_Z)$, $\Delta n = n_o - n_e$. For $n_o > n_e$ ($n_o < n_e$), the material is termed a negative (positive) uniaxial crystals [8, 9]. For materials which are biaxial, the situation is very complex. The three refractive indices $n_X$, $n_Y$ and $n_Z$ along X, Y and Z axis are different. The semiaxes of the ellipsoid can always be chosen in a way that either $n_Z > n_Y > n_X$ or $n_Z < n_Y < n_X$ is fulfilled [4, 2, 6, 7, 3, 5]. By convention, $n_Z > n_Y > n_X$, we can calculate $n_Z - n_Y > n_Y - n_X$ to decide whether the biaxial crystal is positive or negative. If $n_Z - n_Y > n_Y - n_X$, the material is termed as positive in analogy to uniaxial crystals, and if $n_Z - n_Y < n_Y - n_X$, the material is termed as negative in analogy to uniaxial crystals [8, 9]. It is important to note that through refractive indices, the crystal type (positive/negative) is wavelength (temperature) dependent. In a biaxial crystal, the birefringence is taken to be the difference between maximum ($n_Z$) and minimum ($n_X$) refractive index.

**Sellmeier curves and Type-I phase matching angle**: For the phase matching geometry for different crystals, we are interested in the Sellmeier curves along different axes. We have obtained the Sellmeier curves $n_{i=X,Y,Z}^2(\lambda)$ for all the MOFs crystals from

the dielectric tensor $\epsilon_{ij}(\omega)$, computed at wavelength intervals from $\lambda_{min} = hc/E_g$ to $\lambda_{max} = 1100$ nm. $E_g$ is the band gap for different MOFs crystals. We fit the Sellmeier curves for each crystal axis to the standard relation

$$n_i^2(\lambda) = A + \frac{B\lambda^2}{\lambda^2 - C} + \frac{B_1\lambda^2}{\lambda^2 - C_1}, \tag{4}$$

to obtain the Sellmeier coefficients.

For a uniaxial crystal, the Type-I phase-matching angle, $\theta_m$, is taken from the optic axis. The phase-matching "point" in three-dimensional form symmetrical cones [10, 1]. We can obtain $\theta_m$ for negative uniaxial crystal using $\sin^2(\theta_m) = (n_e^2(2\omega)/n_o^2(\omega))(n_o^2(2\omega) - n_o^2(\omega))(n_o^2(2\omega) - n_e^2(2\omega))$, and for positive uniaxial crystal using $\sin^2(\theta_m) = (n_e^2(\omega)/n_o^2(2\omega))(n_o^2(\omega) - n_o^2(2\omega))/(n_o^2(\omega) - n_e^2(\omega))$ [10].

To calculate the phase matching angles $\theta_m$ and $\phi_m$ in biaxial crystals, one must solve the following equation for $\theta$ and $\phi$

$$\frac{k_X^2}{n^{-2} - n_X^{-2}} + \frac{k_Y^2}{n^{-2} - n_Y^{-2}} + \frac{k_Z^2}{n^{-2} - n_Z^{-2}} = 0 \tag{5}$$

where $k_X = \sin\theta\cos\phi$, $k_Y = \sin\theta\sin\phi$, and $k_Z = \cos\theta$. $\theta$ is the angle from the Z-axis and $\phi$ is the angle from the X-axis in the XY plane. $n_X$, $n_Y$ and $n_Z$ are the three principal refractive indices. Equation 5 can be simplified to equation 6 by letting $x = n^{-2}$, $B = [-k_X^2(b+c) - k_Y^2(a+c) - k_Z^2(a+b)]$, $C = k_X^2 bc + k_Y^2 ac + k_Z^2 ab$, where $a = n_X^{-2}$, $a = n_Y^{-2}$, $c = n_Z^{-2}$

$$x^2 + Bx + C = 0 \tag{6}$$

On solving equation 6, we get

$$n = \sqrt{2}/(\sqrt{-B \pm \sqrt{(B^2 - 4C)}}). \tag{7}$$

For biaxial type-I phase matching, we solve

$$\sqrt{2}/(\sqrt{-B_1 - \sqrt{(B_1^2 - 4C_1)}}) = \sqrt{2}/(\sqrt{-B_2 + \sqrt{(B_2^2 - 4C_2)}}), \tag{8}$$

where the subscripts 1 and 2 stand for fundamental and second harmonic wavelengths. On solving Eq. 8, we get $\theta$ and $\phi$ for biaxial type-I phase matching. More details about type-I phase-matching in biaxial crystal are given by Yao and Fahlen [3].

**Glauber Correlation function**: The detection of arrival times of all the photon pairs generated by the MOF crystal is done in a coincidence setup. The probability of detecting two-photon pairs at times $t_1$ and $t_2$ is proportional to the correlation function $G^{(2)}(t_1, t_2)$ given by [11, 12]

$$G^{(2)}(t_1, t_2) = |\langle 0|\left(\hat{E}^{(-)}(t_2)\hat{E}^{(+)}(t_1)\right)|\Psi_2\rangle|^2, \tag{9}$$

where the electric field operators are given by

$$E^{(+)}(r, t) = \frac{1}{\sqrt{V}} \sum_{k,\sigma} e_{k,\sigma}\epsilon_{k,\sigma}\alpha_{k,\sigma}G(\omega)e^{i(k \cdot r - \omega t)} = [E^{(-)}(k, t)]^*, \tag{10}$$

with $\epsilon_{k,\sigma} = \hbar\omega(k,\sigma)/2\epsilon_0 n^2(k,\omega)$. V is the quantization volume, index $\sigma$ is summed over a two-dimensional orthogonal polarization vector. $k$ is an index summed over all wave vectors, $e_{k,\sigma}$ is the two-dimensional unit polarization vector, $\alpha_{k,\sigma}$ is the mode amplitude, $G(\omega)$ is a transmission function that models the detector filter, $\omega$ the frequency, $\epsilon_0$ is the permittivity of free space.

We have assumed that this correlation depends only on a relative time $\tau = t_1 - t_2$. It is possible because we only consider the pump as a gaussian beam and the Hermite-Gauss basis for the generated photons. The right-hand side of Eq. (9) is proportional to the frequency spectrum of the light as

$$G^{(2)}(\tau) \propto |I_0(\tau)|^2, \tag{11}$$

and the light spectrum frequency is given by

$$I_0(\tau) = \int_{-\infty}^{\infty} d\nu \, \Phi\left(\frac{\Delta k L}{2}\right) e^{-\left(\frac{\nu^2}{\sigma^2}\right)} e^{-i\nu\tau}, \tag{12}$$

where $\nu$ is a small detuning frequency, and $\Phi(x)$ is known as the joint spectral function (JSA) which is typically proportional to a "sinc" function. $\Delta k$ is given by

$$\Delta k \approx \left(\frac{n_g}{c}\right)\left(\omega_s - \frac{\omega_p}{2}\right) + \kappa\left(\omega_s - \frac{\omega_p}{2}\right)^2, \tag{13}$$

where $n_g$ is the group velocity difference for the generated photons, and $\kappa$ corresponds to the group velocity dispersion of the generated photon. Since we are interested in Type I degenerate SPDC, the generated photons have the same central frequency that is half the central frequency of the pumped photon. As the generated photons have the same polarization, $n_g$ is and the $\Delta k$ expression becomes

$$\Delta k \approx \frac{\kappa \nu^2}{4}. \tag{14}$$

On combining Eqs. 12 and 14 in Eq. 11, we obtain Glauber Correlation function as

$$G^{(2)}(\tau) \propto \left|\int_{-\infty}^{\infty} d\nu \, \text{sinc}\left(\frac{\kappa \nu^2 L}{4}\right) e^{-\left(\frac{\nu^2}{\sigma^2}\right)} e^{-i\nu\tau}\right|^2. \tag{15}$$

**Counting Rate**: The biphoton generation rate depends on the integral [13]

$$|\Phi(\Delta k_z)|^2 \equiv \left|\int d^3r \left(\chi_{eff}^2(\vec{r}) G_p^*(\vec{r}) g_{\vec{\mu}_1}(x,y) g_{\vec{\mu}_2}(x,y) e^{-i\Delta k_z z}\right)\right|^2 \tag{16}$$

The problem simplifies for the case in which a collimated Gaussian pump beam directed onto an isotropic rectangular crystal (positioned at the origin within a Cartesian coordinate system, with the $\hat{z}$ axis aligned along the optic axis) with dimensions $L_x \times L_y \times L_z$. Assuming our interest lies in gathering the down-converted light through single-mode fibers, only photons generated in the zeroth-order Hermite–Gaussian modes will impact the detected event rate. Consequently, $G_p(\vec{r})$, $g_{\vec{\mu}_1}$, and $g_{\vec{\mu}_2}$ all conform to gaussian functions, leading to the transformation of $|\Phi(\Delta k_z)|^2$ to

$$|\Phi(\Delta k_z)|^2 = C_1 \left|\int_{L_z/2}^{-L_z/2} dz e^{-i\Delta k_z z}\right|^2 \left|\int dx dy \ e^{\left[-(x^2 y^2)\left(\frac{1}{4\sigma_p^2} + \frac{2}{4\sigma_1^2}\right)\right]}\right|, \tag{17}$$

where $C_1 = \left(\frac{\chi^{(2)}_{eff}}{2\pi\sigma_1^2}\right)^2$. We have assumed that both detectors' filters possess identical bandwidths, denoted as $\sigma_1 = \sigma_2$. Expanding the limits of integration over $x$ and $y$ to arbitrarily large values requires only that the transverse width of the crystal exceeds the dimensions of the Gaussian pump beam as well as those of the signal and idler modes. Additionally, in single-mode nonlinear waveguides, light confinement within a considerably smaller beam diameter is achievable without divergence. Under these assumptions, the overlap integral simplifies significantly to [13]

$$|\Phi(\Delta k_z)|^2 = (2\chi^2_{\text{eff}}L_z)^2 \text{sinc}^2\left(\frac{\Delta k_z L_z}{2}\right)\left|\frac{\sigma_p^2}{\sigma_1^2 + 2\sigma_p^2}\right|^2. \tag{18}$$

Presently, an expression can be formulated for the overall down-conversion rate, denoted as $R$, using as a Gaussian pump beam into another Gaussian signal-idler mode, this expression is formulated in relation to the frequencies of the signal and idler, denoted as $\omega_1$ and $\omega_2$ as

$$R = \int d\omega_1 d\omega_2 C_2 \omega_1 \omega_2 \delta(\Delta\omega)\text{sinc}^2\left(\frac{\Delta k_z Lz}{2}\right) \tag{19}$$

Where $C_2 = \frac{|E_p^0|^2(d_{\text{eff}})^2 L_z^2}{2\pi c^2}\frac{n_{g1}n_{g2}}{n_1 n_2}\left|\frac{\sigma_p^2}{\sigma_1^2+2\sigma_p^2}\right|^2$. From energy conservation we have $\omega_p = \omega_1 + \omega_2$. Now we can integrate over one of the frequencies using the delta function. We integrate over the frequency $\omega_2$ getting

$$R = C_2 \int d\omega_1 \omega_1(\omega_p - \omega_1)\text{sinc}^2\left(\frac{\Delta k_z Lz}{2}\right) \tag{20}$$

where the phase matching function is given by

$$\Delta k_z = k(\omega_1) + k(\omega_p - \omega_1) - k(\omega_p). \tag{21}$$

This function is analogous to the expression of equation 13, meaning that in the case of type I SPDC we need to replace $\Delta k$ for this expression.

**Estimation of HOMO-LUMO gap:** To calculate the HOMO-LUMO gap for every ligand in every MOF, first, we retrieve the chemical name of the organic molecules from the chemical name of the MOFs in the Cambridge Structural Database (CSD) [14]. By using conventional inorganic compound naming we were able to filter and obtain the chemical name of the ligand for Zinc based MOFs with one organic molecule as ligand and $Zn^{2+}$ as node. Using OPSIN [15] we converted these names to SMILES and posteriorly with Rdkit [16]. We generated a 3D structure for every molecule and subsequently perform geometry optimization of the coordinates with Dmol3 [17] with Minessota funtional and DNP 4.4 basis set of Dmol3 and obtain the values for highest occupied molecular orbital and the lowest unoccupied molecular orbital, these values are used to estimate the HOMO-LUMO gap. A more refined approximation would be to extract the coordinates of the ligands from the crystal but a correct capping of each ligand is required to represent the correct valence. This chemical name-to-structure approach requires precise chemical naming on the databases, as misspelled or bad labeling is possible.

## 2. Biaxial and Uniaxial MOF properties

Table 1: Crystal system, crystal class, space group, band gap, coordinate transformation, birefringence and crystal type for biaxial crystals and that N is for negative and P is for positive.

| MOFs | Crystal system | Crystal Class | Space group | Band gap (eV) | Transformation | Birefringence (1064 nm) | Crystal type |
|---|---|---|---|---|---|---|---|
| SOMNUG | Triclinic | 1 | P1 | 2.788 | x, y, z -> a, b, c | 0.467967 | N |
| RATDAS01 | Monoclinic | 2 | P21 | 4.3819 | x, y, z -> a, c, b | 0.06553 | P |
| TAKRIJ | Monoclinic | 2 | P21 | 4.4618 | x, y, z -> a, c, b | 0.102171 | P |
| FIPWUY | Monoclinic | 2 | P21 | 4.1972 | x, y, z -> b, a, c | 0.117632 | N |
| LASPZN | Monoclinic | 2 | P21 | 3.6932 | x, y, z -> a, b, c | 0.116657 | P |
| FIRZIP | Monoclinic | 2 | P21 | 3.3416 | x, y, z -> a, c, b | 0.042447 | P |
| PODREH | Monoclinic | 2 | C2 | 2.6703 | x, y, z -> b, c, a | 0.281457 | N |
| RUVBAN | Monoclinic | 2 | C2 | 3.2442 | x, y, z -> c, a, b | 0.325459 | N |
| OFUWIV | Monoclinic | 2 | C2 | 5.0703 | x, y, z -> b, c, a | 0.144579 | P |
| MUYGOE | Monoclinic | 2 | C2 | 3.836 | x, y, z -> b, a, c | 0.09256 | N |
| NELWUY | Monoclinic | 2 | C2 | 4.4495 | x, y, z -> b, c, a | 0.038723 | N |
| AQOROP | Monoclinic | m | PC | 2.3644 | x, y, z -> c, b, a | 0.647674 | N |
| KOMMII | Monoclinic | m | PC | 5.1303 | x, y, z -> a, b, c | 0.04897 | P |
| MOFTIL | Monoclinic | m | PC | 3.2545 | x, y, z -> c, b, a | 0.422821 | N |
| ECIWAO | Monoclinic | m | CC | 2.2834 | x, y, z -> b, a, c | 0.372085 | P |
| IDOGIR | Monoclinic | m | CC | 4.8648 | x, y, z -> c, b, a | 0.085201 | N |
| RUBDOI | Monoclinic | m | CC | 3.4187 | x, y, z -> c, b, a | 0.152212 | N |
| SEYVIC | Monoclinic | m | CC | 4.3136 | x, y, z -> b, c, a | 0.145033 | P |
| ECIHIK | Orthorhombic | 222 | P21212 | 3.8375 | x, y, z -> c, a, b | 0.145033 | P |
| ODIVON | Orthorhombic | 222 | P21212 | 3.429 | x, y, z -> c, a, b | 0.138952 | N |
| WOVPAB | Orthorhombic | 222 | P21212 | 2.5401 | x, y, z -> a, c, b | 0.115771 | N |
| DOYBEY | Orthorhombic | 222 | P212121 | 3.6675 | x, y, z -> c, b, a | 0.111669 | N |
| EWUSIZ | Orthorhombic | 222 | P212121 | 4.5544 | x, y, z -> c, b, a | 0.091731 | P |
| JUKVEQ02 | Orthorhombic | 222 | P212121 | 2.9487 | x, y, z -> c, a, b | 0.080823 | N |
| NIBRUN | Orthorhombic | 222 | P212121 | 3.2788 | x, y, z -> a, b, c | 0.215102 | P |
| NIBSAU | Orthorhombic | 222 | P212121 | 3.277 | x, y, z -> a, b, c | 0.215102 | P |
| VAHWOS01 | Orthorhombic | 222 | P212121 | 3.2556 | x, y, z -> a, b, c | 0.683581 | P |
| WOVPEF | Orthorhombic | 222 | P212121 | 2.9953 | x, y, z -> c, a, b | 0.040637 | N |
| XISBUX | Orthorhombic | 222 | P212121 | 4.9015 | x, y, z -> b, c, a | 0.072898 | N |
| HONCOF | Orthorhombic | mm2 | PCA21 | 3.5635 | x, y, z -> a, b, c | 0.14768 | P |
| DIFKOS | Orthorhombic | mm2 | PCA21 | 5.1694 | x, y, z -> b, a, c | 0.021875 | P |
| FEHJIN | Orthorhombic | mm2 | PCA21 | 3.5443 | x, y, z -> b, c, a | 0.203427 | P |
| VEKNEH | Orthorhombic | mm2 | PNA21 | 3.7591 | x, y, z -> b, a, c | 0.091819 | P |
| IVAZUA01 | Orthorhombic | mm2 | PNA21 | 3.0381 | x, y, z -> a, b, c | 0.227767 | N |
| FIQGEW | Orthorhombic | mm2 | ABA2 | 3.0333 | x, y, z -> a, c, b | 0.318939 | N |
| NUYWUA | Orthorhombic | mm2 | ABA2 | 3.2267 | x, y, z -> b, c, a | 0.166632 | P |
| OXUNUR | Orthorhombic | mm2 | FDD2 | 2.6149 | x, y, z -> c, a, b | 0.31352 | N |

Table 2: Crystal system, crystal class, space group, band gap, coordinate transformation, birefringence and crystal type for uniaxial crystals. npm stands for no phase matching.

| MOFs | Crystal system | Crystal class | space group | Band gap (eV) | Coordinate assignment | Crystal type | Birefringence | $\theta_m$ | $d_{eff}$ (pm/V) |
|---|---|---|---|---|---|---|---|---|---|
| VURQEF | Tetragonal | 422 | P4122 | 3.2709 | X, Y, Z -> a, b, c | N | 0.2916 | 23.67 | 0 |
| ONOCOL01 | Tetragonal | 422 | P41212 | 3.1801 | X, Y, Z -> a, b, c | N | 0.0085 | npm | |
| DIXWAJ | Tetragonal | 422 | P43212 | 3.3833 | X, Y, Z -> a, b, c | N | 0.0588 | 42.19 | 0 |
| EGIKUB02 | Tetragonal | 422 | P43212 | 3.3028 | X, Y, Z -> a, b, c | P | -0.2115 | 27.78 | 0 |
| HICDUV | Tetragonal | 422 | P43212 | 2.6224 | X, Y, Z -> a, b, c | P | -0.0826 | npm | |
| IMIDZB01 | Tetragonal | 4mm | I41CD | 4.6497 | X, Y, Z -> a, b, c | P | -0.0253 | npm | |
| JETVOW | Hexagonal | 32 | P3121 | 2.5202 | X, Y, Z -> a, b, c | P | -0.0454 | npm | |
| SAQFUM | Tetragonal | -42m | I-42D | 3.5932 | X, Y, Z -> a, b, c | N | 0.0924 | 41.91 | 0.1 |
| HUTZAX01 | Hexagonal | 32 | P3221 | 3.0568 | X, Y, Z -> a, b, c | P | -0.1065 | 48.33 | 1.18 |
| QAMFUF01 | Hexagonal | 32 | P3221 | 2.3881 | X, Y, Z -> a, b, c | P | -0.0355 | npm | |
| BEKVOD | Trigonal | -3m1 | R3C | 3.5104 | X, Y, Z -> a, b, c | P | -0.0095 | npm | |
| WUJFOW | Hexagonal | 6 | P61 | 4.7085 | X, Y, Z -> a, b, c | N | 0.0266 | 59.76 | 0.11 |

## 3. Biaxial crystal Phase-matching angle (θ, φ)

### 3.1. Test Case BiBO and KTP

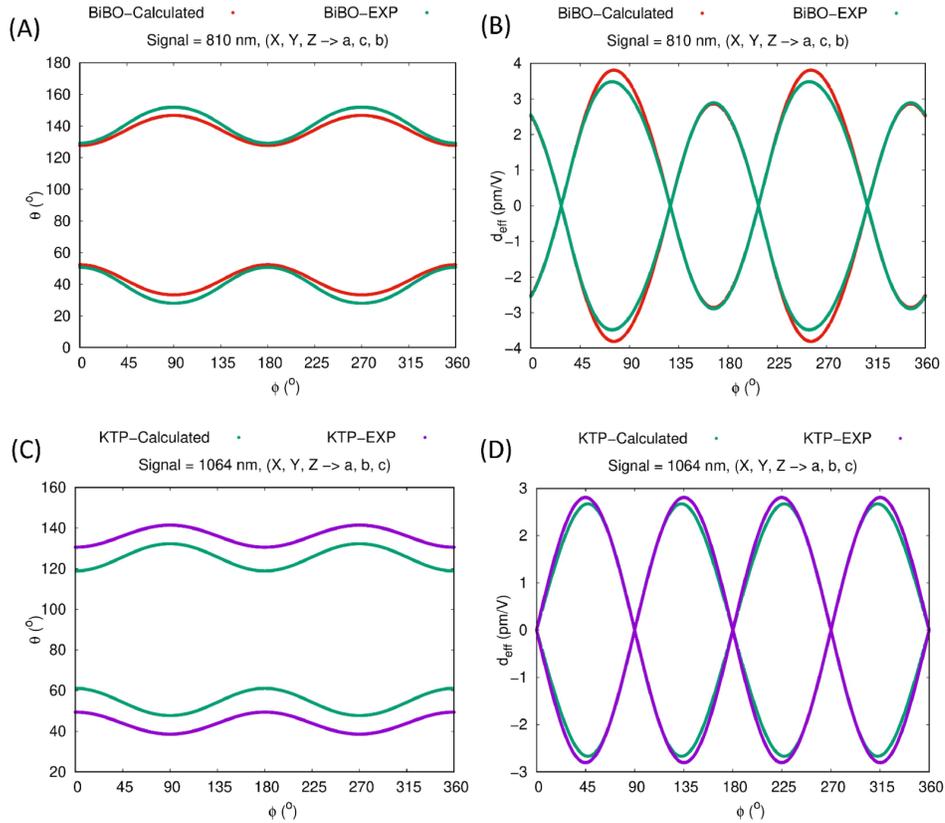

**Figure 1.** (A) Experimental [5] and calculated phase-matching angles for BiBO crystal belonging to crystal class 2. (B) $d_{\text{eff}}$ of BiBO crystal belonging to crystal as a function of phase matching angle, $\theta$. (C) Experimental [3] and calculated phase-matching angles for KTP crystal belonging to crystal class mm2. (B) $d_{\text{eff}}$ of KTP crystal belonging to KTP crystal as a function of phase matching angle, $\phi.\theta$ is not shown for clarity

## 3.2. Biaxial crystals used in our study

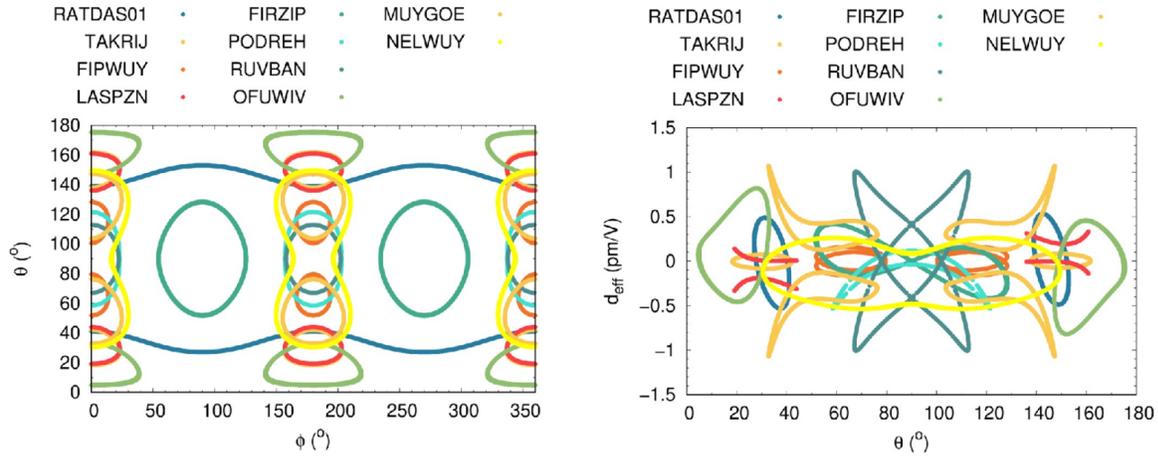

**Figure 2.** (Left) Phase-matching angle for MOF crystal belonging to crystal class 2. (Right) $d_{\text{eff}}$ of crystals belonging to crystal class 2 as a function of phase matching angle, $\theta$. $\phi$ is not shown for clarity

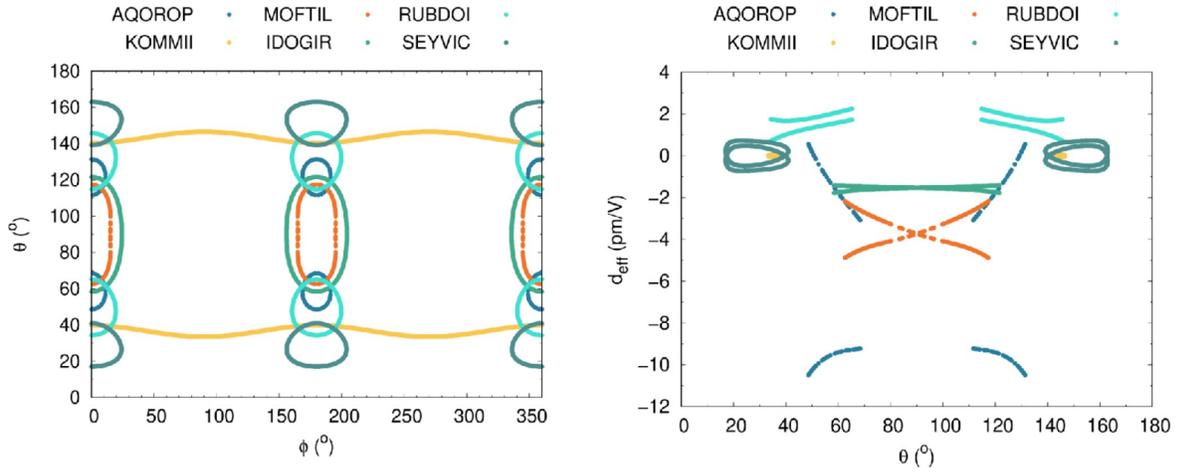

**Figure 3.** (Left) Phase-matching angle for MOF crystal belonging to crystal class m. (Right) $d_{\text{eff}}$ of crystals belonging to crystal class m as a function of phase matching angle, $\theta$. $\phi$ is not shown for clarity

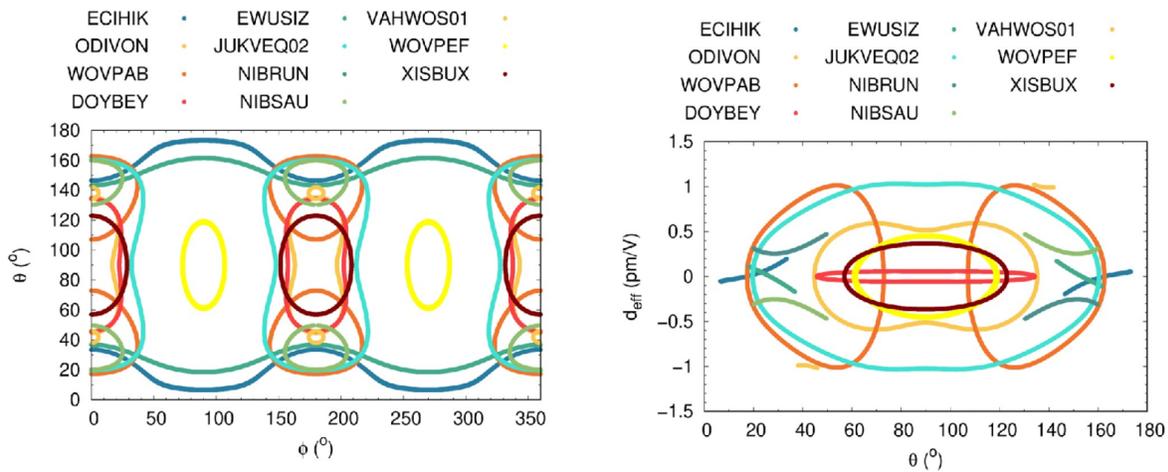

**Figure 4.** (Left) Phase-matching angle for MOF crystal belonging to crystal class 222. (Right) $d_{]rmeff}$ of crystals belonging to crystal class 222 as a function of phase matching angle, $\theta$. $\phi$ is not shown for clarity

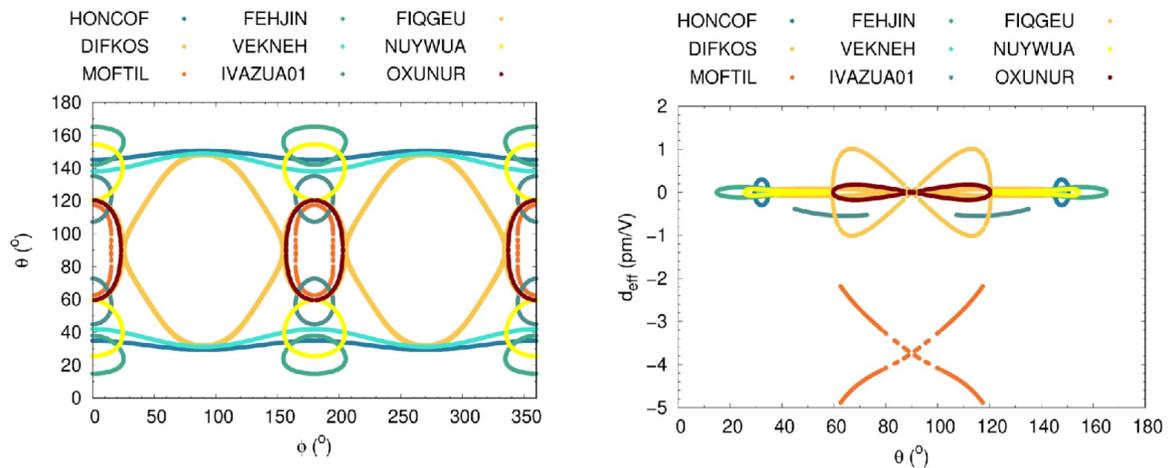

**Figure 5.** (Left) Phase-matching angle for MOF crystal belonging to crystal class mm2. (Right) $d_{\text{eff}}$ of crystals belonging to crystal class mm2 as a function of phase matching angle, $\theta$. $\phi$ is not shown for clarity

## 4. Second-order nonlinear susceptibility tensor ($\chi^{(2)}$)

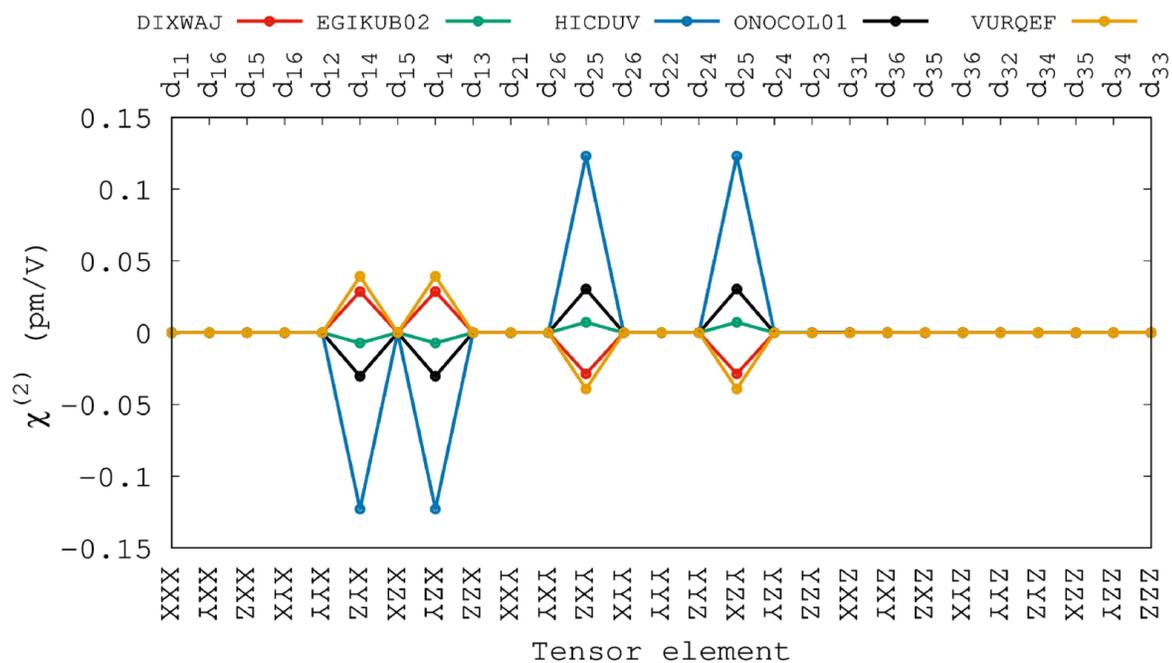

**Figure 6.** $\chi^{(2)}$ of different MOF crystal belonging to crystal class 422.

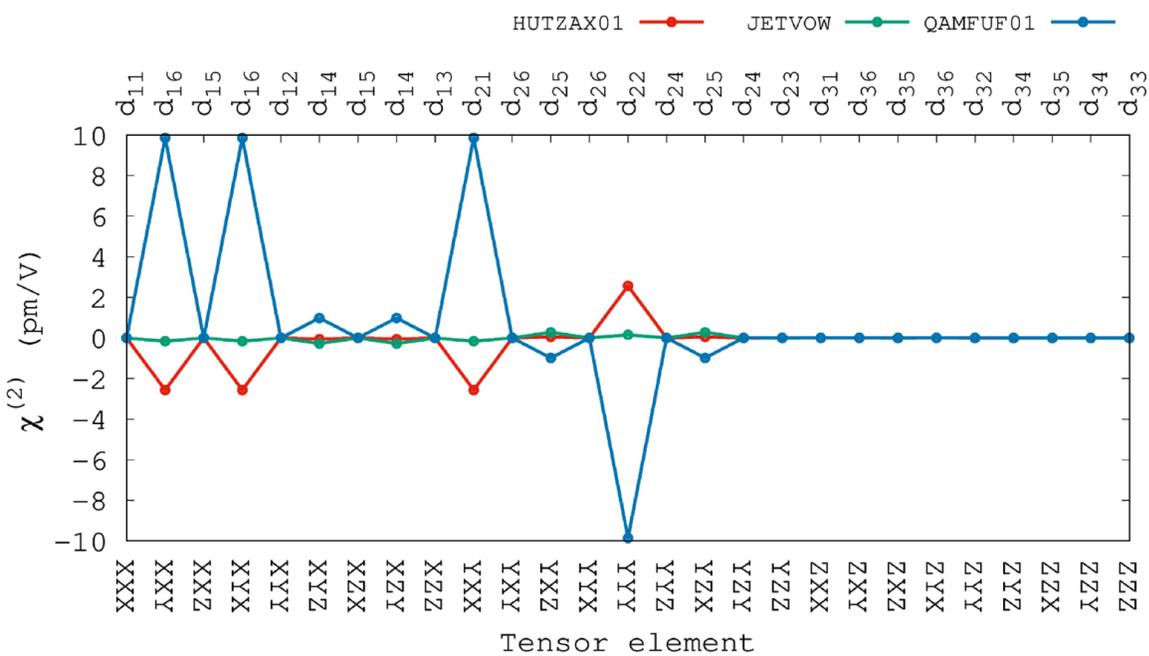

**Figure 7.** $\chi^{(2)}$ of different MOF crystal belonging to crystal class 32.

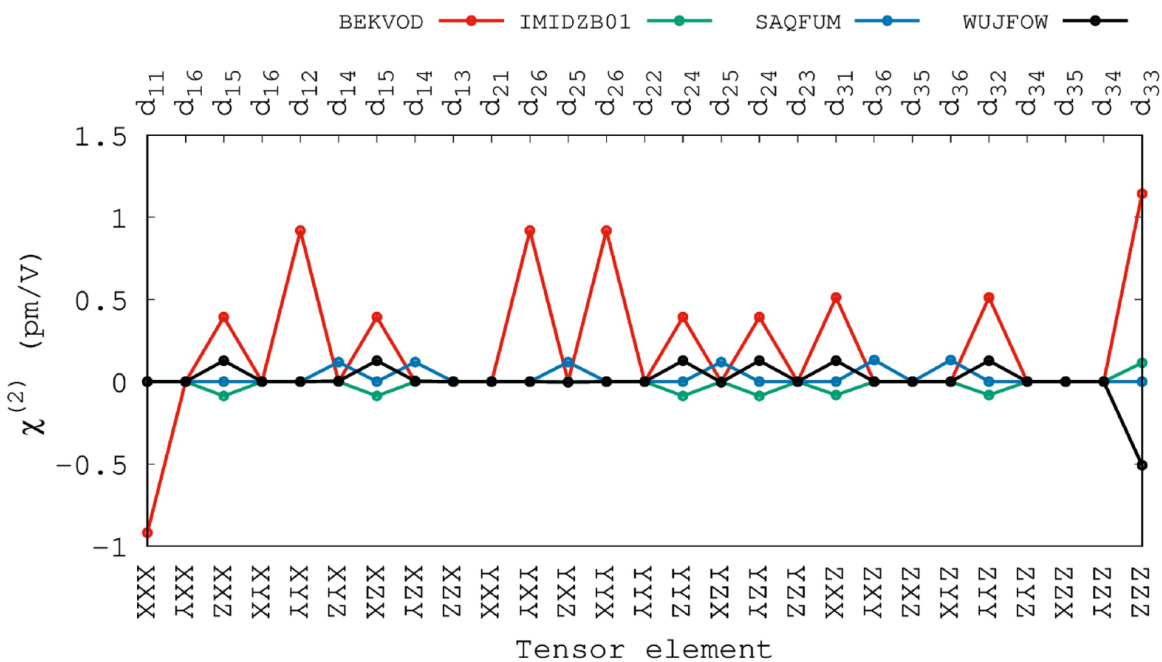

**Figure 8.** $\chi^{(2)}$ of different MOF crystal belonging to crystal class -3m1, 4mm, -42m and 6.

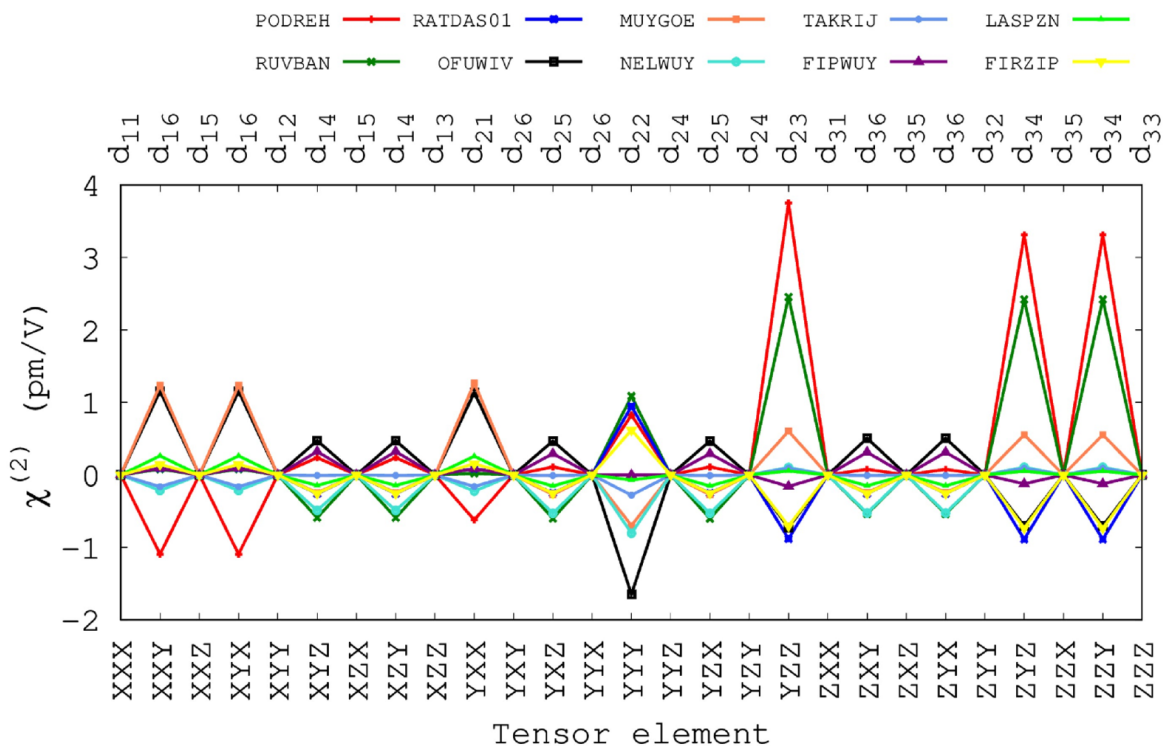

**Figure 9.** $\chi^{(2)}$ of different MOF crystal belonging to crystal class 2.

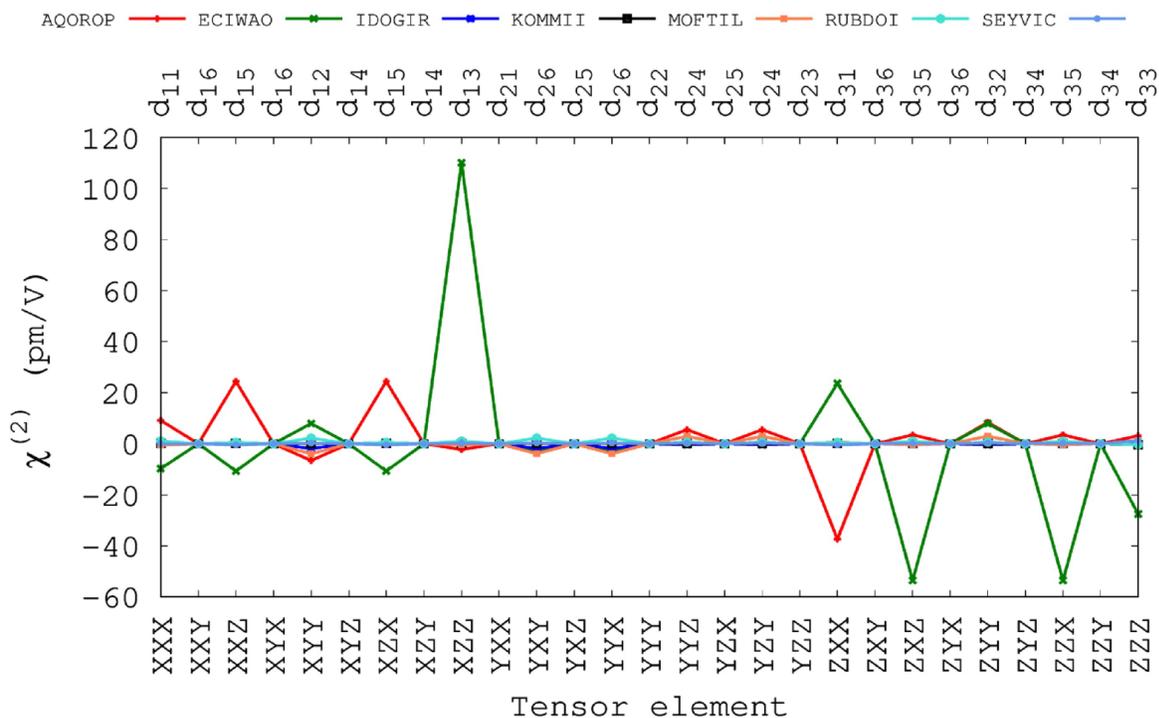

**Figure 10.** $\chi^{(2)}$ of different MOF crystal belonging to crystal class m.

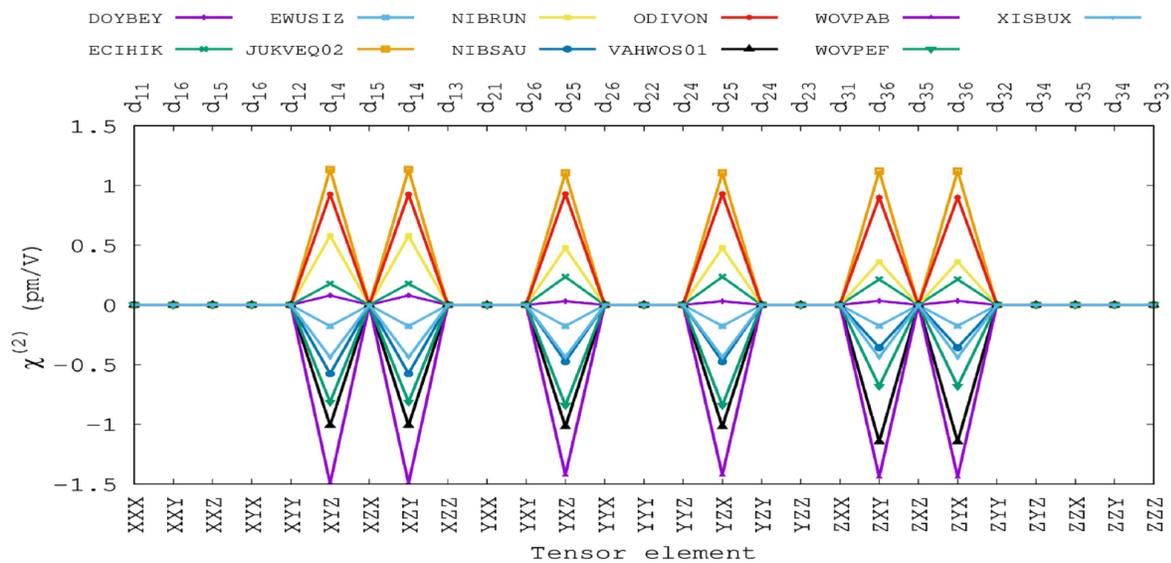

**Figure 11.** $\chi^{(2)}$ of different MOF crystal belonging to crystal class 222.

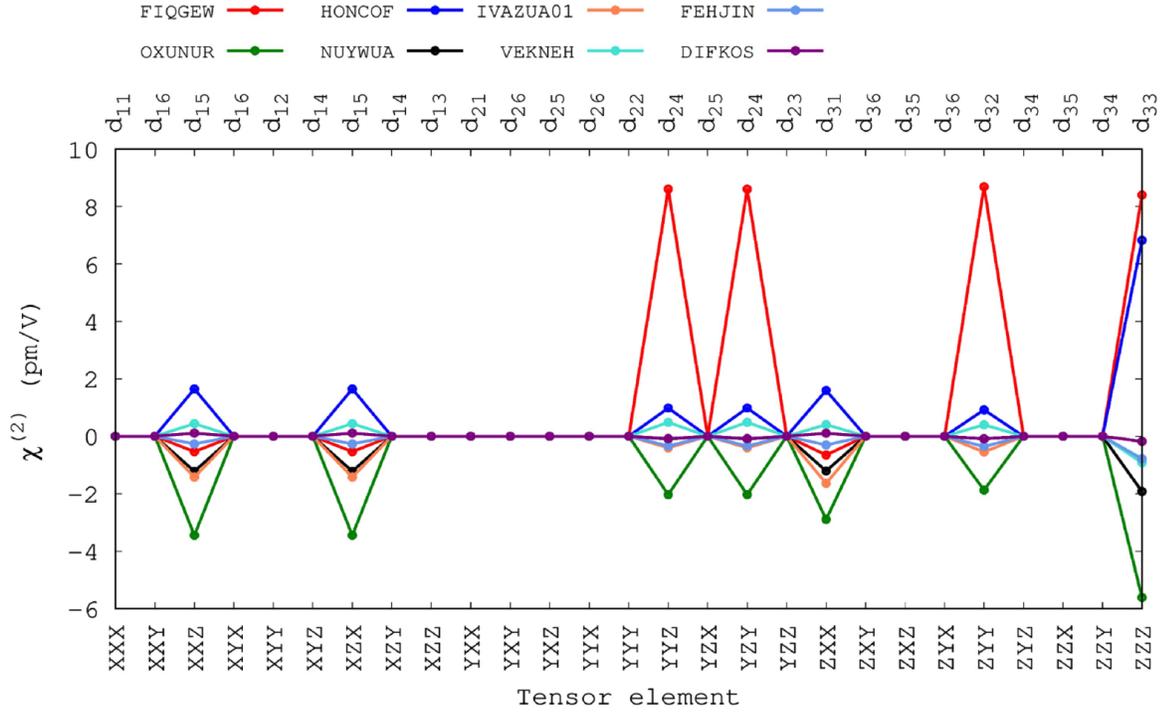

**Figure 12.** $\chi^{(2)}$ of different MOF crystal belonging to crystal class mm2.

## 5. Dispersive properties of nonlinear crystals

We have used the following functional form of sellmeier equation to fit the refractive indices values.

$$n_i^2(\lambda) = A + \frac{B}{\lambda^2 - C} - B_1\lambda^2 \tag{22}$$

Here, $\lambda$ is in nm. We used the fitted sellmeier equations to obtain refractive indices values at different wavelength and then estimate the GVM (see Eq. 23).

**Group velocity mismatch (GVM)**: The GVM for the degenerate case for uniaxial and biaxial MOF crystal was calculate using [18, 12]

$$D_+(GVM) = \frac{1}{u_o(\Omega_p/2)} - \frac{1}{u_p(\Omega_p)}, \tag{23}$$

where, $u(\Omega)$ are the group velocities at the frequency $\Omega$. For uniaxial crystals, phase matching equations are iterated to calculate the phase matching angles at a given wavlength and then used Eq. 23 to estimate the GVM. For biaxial crystals [3], we estimated the gvm at all possible phase matching angles (see SI section 1) at a given wavelength and then estimated the GVM using Eq. 23. We picked the minimum value from all the absolute values of GVM. Figure 13 shows the GVM of KDP, BBO and BiBO using the experimental data [19, 20, 21] and are shown in Fig. 13A as KDP-EXP, BBO-EXP and BIBO-EXP. We have also included the calculated GVM of BiBO as BIBO-CAL in Fig. 13.

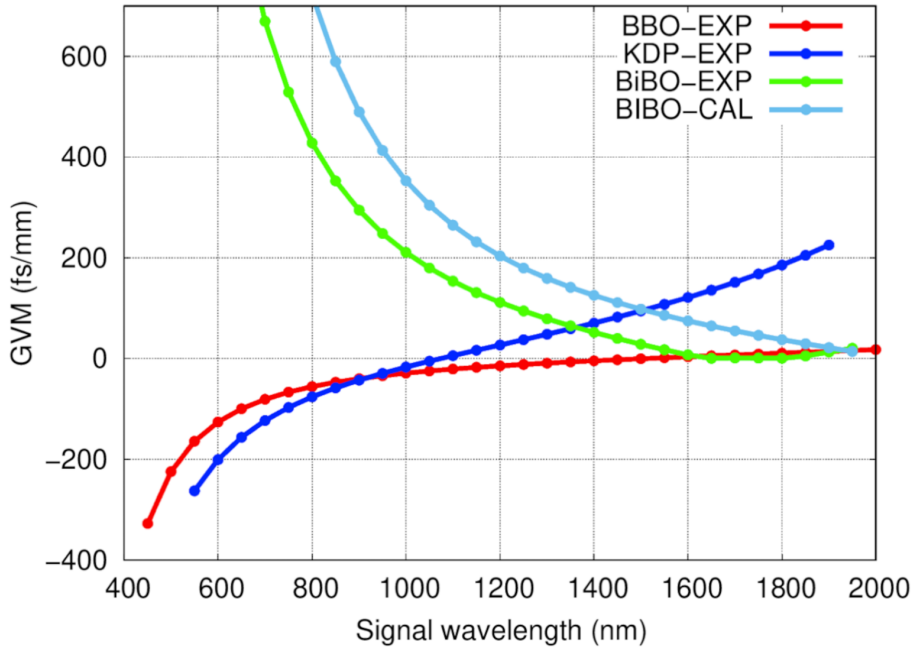

**Figure 13.** Calculated GVM for KDP, BBO and BiBO using experimental and calculated refractive indices.

**Group velocity dispersion(GVD)**: The GVD for biaxial and uniaxial using $\text{GVD} = \frac{\lambda^3}{2\pi c^2} \frac{d^2 n}{d\lambda^2}$, with appropriate signal depending on whether a given crystal is positive uniaxial/biaxial or negative uniaxial/biaxial (see SI section 1 and method section in the main text).

**Phase matching function**: Taking into account that the conditions of perfect phase matching are fulfilled and the crystal is uniaxial, we can calculate the spectral acceptance bandwidth of the phenomenon at a fixed pump wavelength $\lambda_p$ using the joint spectral function, which depends on the phase matching mismatch $\Delta k$:

$$\Phi \propto \text{sinc}^2\left(\frac{\Delta k L}{2}\right) \quad (24)$$

where $L$ is the length of the crystal. The phase mismatch can be calculated directly using a relation between the refractive indices of the pumped and the generated photons:

$$\Delta k = (n_o(\lambda_s) - n_{\text{eff}}(\lambda_p, \theta)) \frac{4\pi}{\lambda_p} \quad (25)$$

where $n_o$ and $n_{\text{eff}}$ are the ordinary and effective refractive indices, respectively. We assume the crystal is positively uniaxial. For a negatively uniaxial crystal, the ordinary refractive index is calculated using the output wavelength ($\lambda_s$), and the effective refractive index is calculated using the pumped wavelength ($\lambda_p$) for SPDC (Spontaneous Parametric Down-Conversion). Consequently, the spectral acceptance bandwidth can be determined by estimating the full-width at half maximum (FWHM) of the normalized joint spectral function (24).

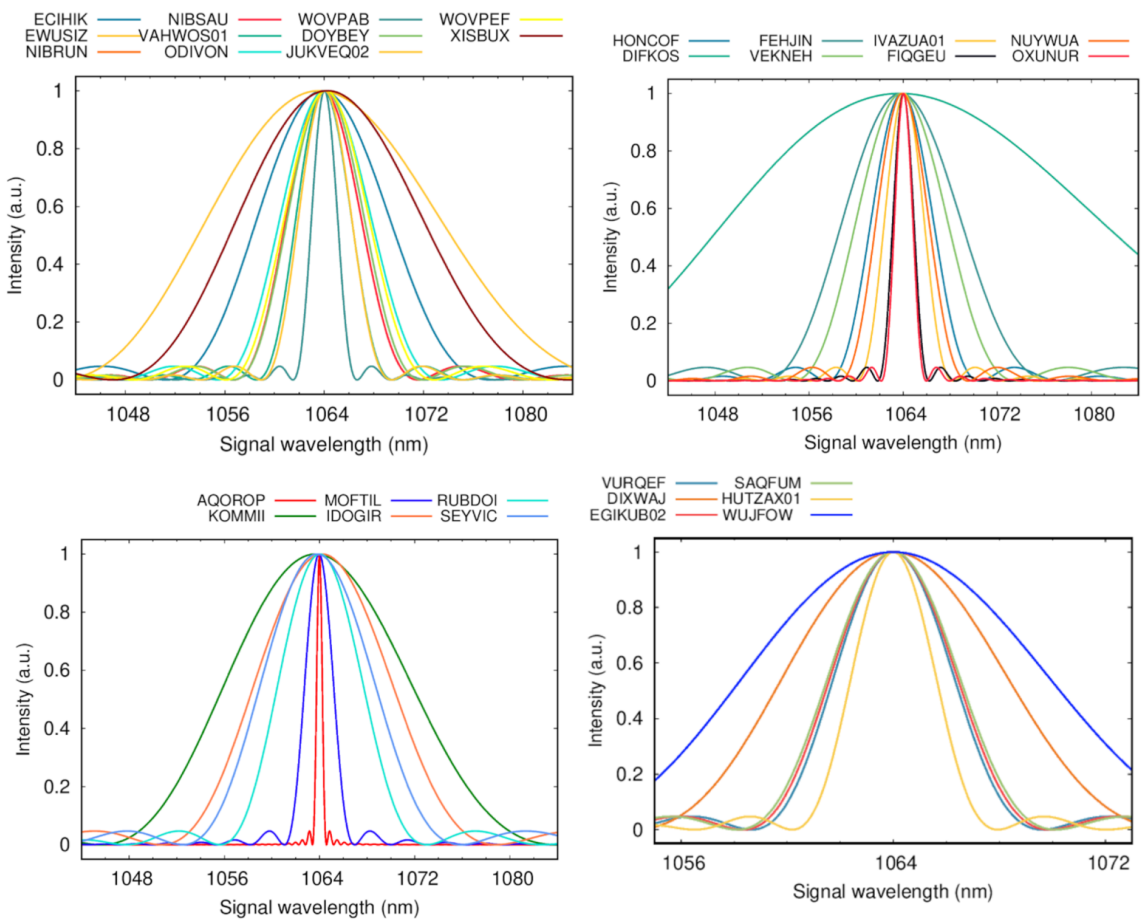

**Figure 14.** Spectral acceptance bandwidth for different crystal class. (Top left) Crystals belonging to crystal class 222. (Top right) Crystals belonging to crystal class mm2. (Bottom left) Crystals belonging to crystal class m. (Bottom right) Uniaxial crystals.



    \end{document}